\documentstyle[aps,pre,preprint,floats,eqsecnum,tighten,epsfig]{revtex}
\preprint{NA/MS/LP/96011}
\draft
\begin{document}
\title{Energy flow, partial equilibration and 
effective temperatures in systems with slow dynamics}
\author{Leticia F. Cugliandolo\cite{add1}}
\address{Laboratoire de Physique Th\'eorique des Liquides, Jussieu\\
4, Place Jussieu,
F-75005 Paris France.}
\author{Jorge Kurchan\cite{add2}}
\address{\'Ecole Normale Sup\'erieure de Lyon, 
46, All\'ee d'Italie, F-69364 Lyon Cedex 07, France.}
\author{Luca Peliti\cite{add3}}
\address{Groupe de Physico-Chimie Th\'eorique\cite{add4}
\\
ESPCI, 10, rue Vauquelin, F-75231 Paris Cedex 05, France,  \\
and
\\
Dipartimento di Scienze Fisiche, Unit\`a INFM, \\
Universit\`a ``Federico II'', Mostra d'Oltremare, Pad.~19,
I-80125 Napoli, Italy.}
\date\today
\maketitle
\begin{abstract}
We show that, in nonequilibrium 
systems with small heat flows, 
 there is a time-scale dependent effective temperature
which plays the same role as the thermodynamical temperature,
 in that it controls the direction of heat flows
and acts as a criterion for thermalization.
We simultaneously treat the case of stationary systems with 
weak stirring and of glassy systems that age after cooling and show that 
they exhibit very similar behavior provided that time dependences are 
expressed in terms of the correlations of the system.
We substantiate our claims with
examples taken from solvable models with non-trivial
low-temperature dynamics, 
but argue that they have a much wider range of validity.
We suggest experimental checks of these ideas.
\end{abstract}

\section{Introduction}
\label{Introduction}
No physical system is ever in thermodynamical equilibrium. When
we apply thermodynamics or statistical mechanics, we idealize the
situation by assuming that ``fast'' processes have taken place, and ``slow''
ones will not: hence, we define an observation time scale 
which distinguishes these two kinds of processes~\cite{Ma}. It follows
that the same system can be at equilibrium on one scale,
and out of equilibrium on another, and, more strikingly, that it can
be at equilibrium, but exhibiting different properties, on two
scales at once.

Since the assumption of thermal equilibrium lies at the heart of 
statistical mechanics, it is usually hard to make these considerations
without a strong appeal to one's intuition. We show in the following
that they can be made in fact quite precise for a class of systems
characterized by very slow energy flows. These systems are out of
equilibrium, either because they are  very gently ``stirred'',
i.e., work is constantly made on them, or because they
have undergone a quench from higher temperatures a long time ago.

 The most typical example of such a system is a  piece of glass  that 
has been in a room at constant temperature for several months.
Since the glass itself is not in equilibrium, 
 we have in principle 
no right of talking about ``the temperature of the glass'', but only
about the temperature of the room. However, we may 
legitimately ask what temperature would indicate
 a  thermometer 
brought into contact with the glass and, again, we would be very surprised
if it did not coincide with the room temperature.
We would be even more surprised if, putting two points of the
glass in contact with both ends of a copper wire, a heat 
flow were established through it.

In other words, although we know that  equilibrium thermodynamics
does not apply for the glass, 
we implicitly assume that some concepts  that apply  for equilibrium  are
still relevant for it. This is not because the glass is ``near 
equilibrium'' but rather because 
it has been relaxing for a long time, and therefore thermal  
flows are small.

Many attempts have been done to extend the concepts of
thermodynamics to nonequilibrium systems---such as systems
exhibiting spatiotemporal chaos or weak turbulence\cite{Krch,Hosh}.
In this context, Hohenberg and 
Shraiman \cite{Hosh}  have defined  an effective ``temperature''
for  stationary nonequilibrium systems
through  an expression involving the response, 
the correlation and the temperature of the bath.
A closely related expression appears naturally in the theory
of nonstationary systems exhibiting aging \cite{Cuku1,Cuku2}, such as glasses.

We show here that this expression indeed deserves the name of temperature,
because
(i) the effective temperature associated with a time scale is the one 
measured on the system by a thermometer, in contact with the system,
whose reaction time is equal to the time scale, 
(ii) it determines the direction of heat flows
within a time scale and  (iii) it acts as a criterion for thermalization.

We shall here consider simultaneously two different conditions 
in which a  regime with small flows of energy exists:
\begin{enumerate}
\item Ordinary thermodynamical systems in contact with a heat bath
at temperature $T$ that are
slowly driven (``stirred'') mechanically. The driving force is proportional
to a small number which we shall denote $D$.

 The observation time and the manner of the stirring are such that, 
for long enough times, the system enters a {\em stationary}, 
time-translational invariant (TTI) regime \cite{Foot0}:
one-time average quantities are independent of time, two-time quantities
depend only upon time-differences, etc.  Stationarity is a weaker condition than thermodynamical
equilibrium, since it implies loss of memory of the initial condition but not
all other properties that are linked with the Gibbs-Boltzmann distribution.
\item  Purely relaxational systems 
that have been prepared through some cooling procedure ending
at time $t=0$ and are kept in contact with a heat bath at a constant 
temperature $T$ up to a (long) waiting time $t_{\rm w}$ (as in the example of
the glass). $t_{\rm w}$ is also usually called ``annealing time'' in the glass 
literature.
 In this case physical quantities need not be TTI,
 and in interesting cases they will
keep a dependence upon $t_{\rm w}$ (and also, in many cases, upon the cooling procedure)
 for all later times $t=\tau+t_{\rm w}$.
  
\end{enumerate}

We shall treat in parallel the ``weak stirring'' ($D \rightarrow 0$)  and
the ``old age'' ($t_{\rm w} \rightarrow \infty$) limits: both taken
{\em after the thermodynamical limit of infinite number of degrees of freedom}.
We show that they lead to the same behavior, from the point of view of thermalization
and effective temperatures, provided that one expresses time 
dependences in terms of
the correlations of the system \cite{Cukulepe}. 

As a test of our ideas, we discuss thermalization in the context of the mean field
theory of disordered systems, or the low-temperature generalization of
the Mode Coupling Equations, but the nature of our results makes us confident
that they have a much wider range of validity.

In Section II 
we recall the generalization of the fluctuation-dissipation relation to 
the nonequilibrium case.
In Section III we consider the reading of 
a thermometer coupled to a system: when the system is in equilibrium we
show that the thermometer measures the temperature of the heat bath, while when 
it is out of equilibrium it measures different effective temperatures
depending on the observation time scale. These effective temperatures are equal or 
higher than the one of the bath and are closely related to the FDT violation factor 
\cite{FootSo,Cuku1,Cuku2} introduced 
to describe the out-of-equilibrium dynamics of glassy systems.
In Section IV we recall  how  time scales or correlation scales are defined
in systems with slow dynamics. We then argue that, if the FDT violation factor
is well defined within a time scale, a single degree of freedom thermalizes
within that time scale to the corresponding effective temperature.
In Section V we extend this analysis to several degrees of freedom and
show that the effective temperature determines 
the direction of heat flows, and can be used as a thermalization criterion.
In Section VI we discuss various phenomenological ``fictive temperature'' ideas 
that have been used for a long time in the theory of structural glasses.
Our conclusions are summarized in Section VII, where some experimental implications
of our work are suggested.

\section{The fluctuation-dissipation relation out of equilibrium}
Let us consider a system with $N$ degrees of freedom 
$(s_1,...,s_N)$, whose
dynamics is described by Langevin equations of the form
\begin{equation}
\dot s_i=b_i(s)+\eta_i(t)\; ,
\end{equation}
where $\eta_i(t)$ is the Gaussian thermal noise. 
For {\em unstirred\/} systems, we consider purely
relaxational dynamics, where the average velocity $b_i(s)$
is proportional to the gradient of the Hamiltonian $E(s)$:
\begin{equation}
b_i(s)=-\sum_j\Gamma_{ij}\frac{\partial E(s)}{\partial s_j}
\; .
\end{equation}
The symmetric matrix $\Gamma$ is related to the
correlation function of the noise $\eta$ by the Einstein relations
\begin{equation}
\langle \eta_i(t)\eta_j(t') \rangle =2T\Gamma_{ij}\delta(t-t')
\; ,
\end{equation}
where $T$ is the temperature of the heat bath.
Averages over the thermal history,
i.e., averages over many realizations of the same experiment 
 with different realizations of the heat bath, will be denoted
 by angular brackets.
We assume of course $\langle \eta_i(t) \rangle =0$, $\forall i,t$. We have
chosen the temperature units so that Boltzmann's constant is equal
to 1. The equilibrium distribution is then proportional
to the Boltzmann factor $\exp(-E/T)$.

For {\em stirred\/} systems, we add to $b_i$ a perturbation
proportional to $D$, that cannot
be represented as the gradient of a function (i.e., is not
purely relaxational), but is otherwise generic.
We then have $W\equiv \langle \sum_i b_i(s)\dot s_i \rangle >0$
at stationarity, meaning that work is being made on the system \cite{FootHo}.

We denote the observables (energy, density, magnetization, etc.) by $O(s)$.
 Throughout this work we shall denote by $t_{\rm w}$ or $t'$ 
the earliest time (to be related to the waiting time),  $t$ the latest 
time, and $\tau$ the relative time $t-t_{\rm w}$. These times are measured,
in the case of the unstirred systems which exhibit aging, from the
end of the cooling procedure.

Given two observables $O_1$ and $O_2$, we define their correlations 
$C_{12}(t,t_{\rm w}) \equiv  \langle O_1(t)O_2(t_{\rm w})\rangle - \langle O_1(t)\rangle 
\langle O_2(t_{\rm w})\rangle $, and their mutual response 
\begin{equation}
R_{12}(t,t_{\rm w}) \equiv \frac{\delta O_1(t)}{\delta h_2(t_{\rm w})}\;,
\end{equation}
where $h_2$ appears in a perturbation of the
Hamiltonian of the form $E\to E-h_2(t)O_2$.
Obviously, causality implies $R_{12}(t,t_{\rm w})=0$ for $t< t_{\rm w}$.
It is also useful to introduce the integrated response (susceptibility)
\begin{equation}
\chi_{12}(t,t_{\rm w}) \equiv \int_{t_{\rm w}}^{t} dt' \, R_{12}(t,t')
\; .
\label{m}
\end{equation}
Let us now make a parametric plot~\cite{Cuku2}
of $\chi(t,t_{\rm w})$ vs.\ $C(t,t_{\rm w})$ for several increasing values of $t_{\rm w}$.
We thus obtain a limit curve $\lim_{t_{\rm w}\to\infty}  \chi(t,t_{\rm w})=\chi(C)$.
In the case of a weakly driven system, we wait for stationarity and plot
$\chi(t-t_{\rm w})$ vs.\ $C(t-t_{\rm w})$ for several decreasing values
of the driving $D$.
We thus obtain the curve $\lim_{D\to 0} \chi(t-t_{\rm w},D)=\chi(C)$.

The fluctuation-dissipation theorem
(FDT)  relates the response and correlation function at equilibrium. One has
\begin{mathletters}
\label{FDT1}
\begin{eqnarray}
R_{12}(t-t_{\rm w}) &=& \frac{1}{T} \; \frac{\partial   
C_{12}(t-t_{\rm w})}{\partial t_{\rm w}} 
\; ,
\\
\chi_{12}(t-t_{\rm w}) &=& \frac{1}{T} \; (C_{12}(0)-C_{12}(t-t_{\rm w}))
\; .
\end{eqnarray}
\end{mathletters}
If the equilibrium distribution is asymptotically reached
for $t_{\rm w}\to \infty$ (or $D\to 0$ in the case of strirred systems)
the FDT implies that the limit curve $\chi(C)$ is a straight line of
slope $-1/T$.

\begin{figure}
\centerline{\epsfxsize=10cm
\epsffile{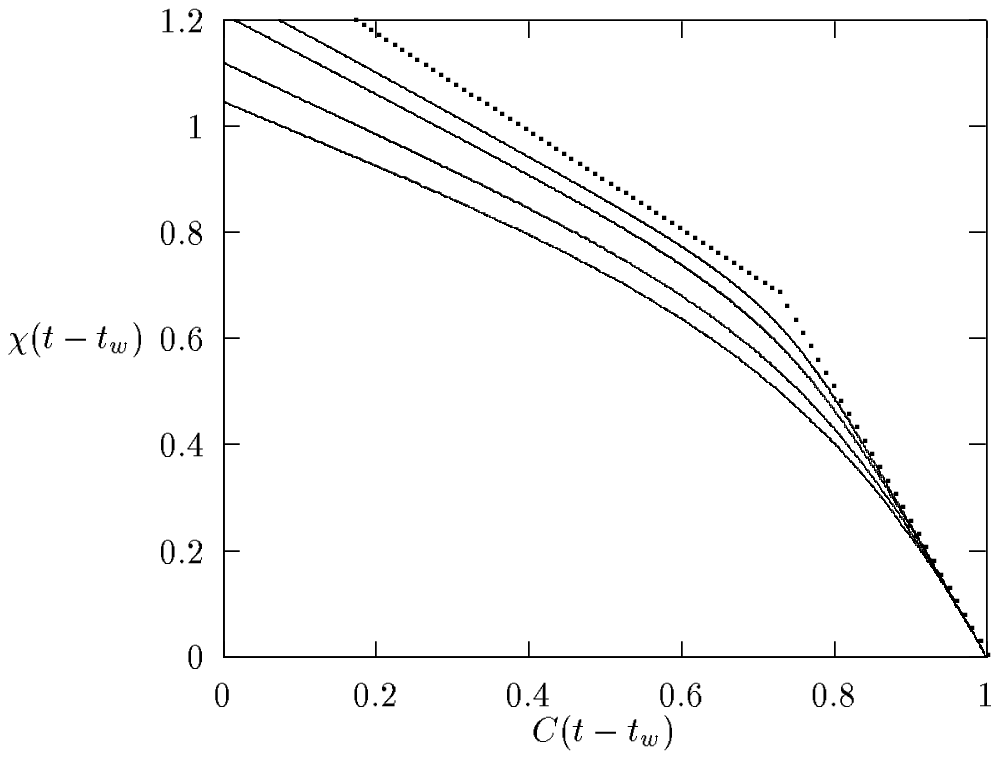}
}
\caption{The susceptibility $\chi(t-t_{\rm w})$ vs.\ 
the auto-correlation function $C(t-t_{\rm w})$
for the model of Appendix \ref{APPmodel} once stationarity is achieved. 
The parameter $D$ is equal to 0.05, 0.375, 0.025, 0.0125
respectively, from bottom to top. The dots represent the analytical 
solution for the limit $D\to 0$. One sees that, in this limit, the 
FDT violation factor $X(C)$ tends {\em continuously\/} to the 
dotted straight lines. 
The value of $C$ at the breakpoint is $C^{EA}$, the Edwards-Anderson
order parameter or the ergodicity breaking parameter in the language of 
the MCT.}
\label{FIGp3chialpha}
\end{figure}

\begin{figure}
\centerline{\epsfxsize=10cm
\epsffile{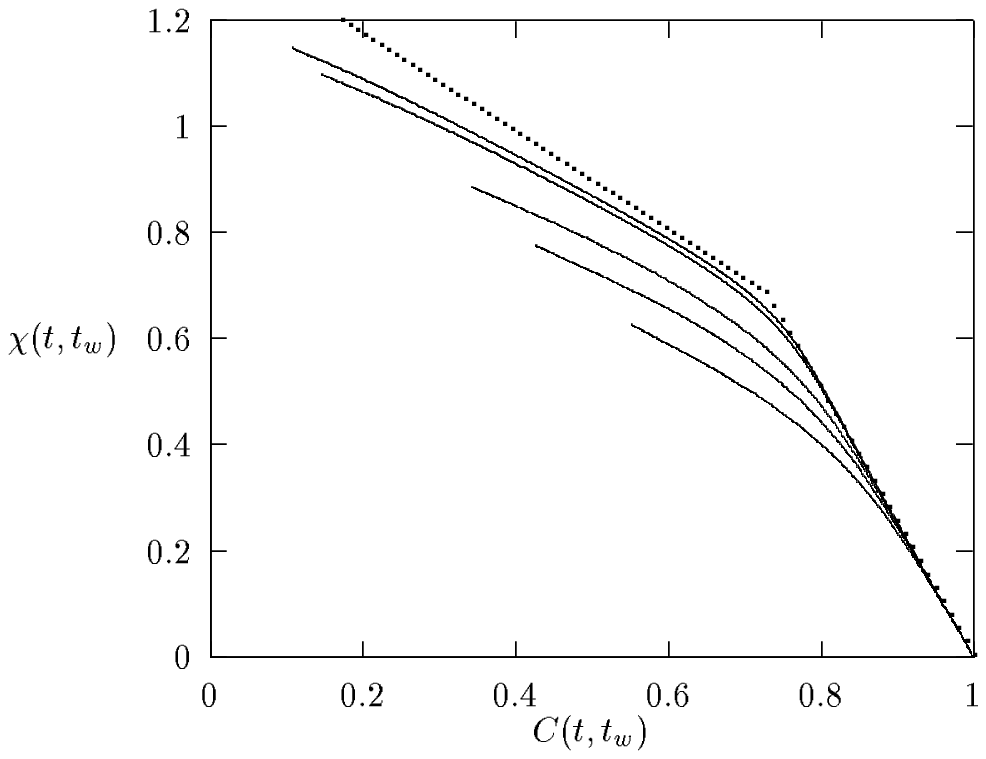}
}
\caption{The susceptibility $\chi(t,t_{\rm w})$ vs.\ the auto-correlation 
function $C(t,t_{\rm w})$
for the model of Appendix \ref{APPmodel}
 at $T<T_{\rm g}$ ($D$ is strictly zero).  The full curves 
correspond to different total times $t$,  equal, from bottom to top, 
to 12.5, 25, 37.5, 50,75 respectively ($t_{\rm w} > t/4$ throughout).
The dots represent the analytical solution  when $t_{\rm w} \to \infty$. Neither
 $\chi(t,t_{\rm w})$ nor $C(t,t_{\rm w})$ achieve stationarity (see Fig.~\ref{FIGp3ctwa1} below). 
\label{FIGp3chitfinite}
}
\end{figure}

However, there is a family of systems for which the limiting curve
$\chi(C)$ {\it does not\/} approach a straight line (see Figs.~\ref{FIGp3chialpha} and \ref{FIGp3chitfinite}).
For driven systems this means that the slightest stirring 
is sufficient to produce a large departure from   
equilibrium {\em even at stationarity\/} (Fig.~\ref{FIGp3chialpha}), while in the case of a  
relaxational system it means that the system 
is unable to equilibrate within experimental times (Fig.~\ref{FIGp3chitfinite}).

Let us denote by  $-X(C)/T$ the slope of the curve $\chi(C)$:
\begin{equation}
\frac{d \chi(C)}{d C} \equiv  -\frac{X(C)}{T} \;.
\end{equation}
This corresponds to
\begin{equation}
R(t,t_{\rm w})=\frac{X(C)}{T} \; \frac{\partial C(t,t_{\rm w})}{\partial t_{\rm w}} \;,
\label{QFDT}
\end{equation}
where the derivative is taken with respect to the earlier time.
We have thus defined $X(C)$, the FDT violation factor 
\cite{Cuku1,Cuku2}, for nonequilibrium systems with slow dynamics.

When FDT holds, for $D\to 0$ (or $t\ge t_{\rm w}\to\infty$
respectively)  we can treat the system as being in equilibrium and
$X$ tends to 1 in the limit.
When this does not happen, we may inquire about the physical meaning of $X(C)$.
In order to answer this question, let us first recall the relationship
there is between the FDT and the equipartition of energy.

\section{Frequency dependent thermometers that measure effective temperatures}
 We use a harmonic oscillator of frequency $\omega_o$ 
 to measure the ``temperature'' of a
 degree of freedom $O(s)$ ($O(s)$ may be 
 the energy, or some spatial Fourier component of
the magnetization). At the waiting time $t_{\rm w}$ we weakly couple 
the oscillator to the system via $O(s)$,
 while we keep the system in contact with a heat bath at temperature $T$.
  We wait for a short time until the average energy of the oscillator
has stabilized.
 If the system were in equilibrium, 
by the principle of equipartition of energy  
we would have $\langle E_{\rm osc}\rangle =T$.

 Assuming linear coupling, the hamiltonian reads:
\begin{equation}
E_{\rm total}= E(s) + E_{\rm osc} + E_{\rm int}
\; ,
\end{equation}
where
\begin{eqnarray}
E_{\rm osc} &=& \frac{1}{2} {\dot x}^2 + \frac1{2} \, \omega_o^2 x^2
\; ,
\\
E_{\rm int} &=& -a O(s) x
\; .
\end{eqnarray}
The equation of motion of the oscillator reads
\begin{equation}
{\ddot x}= - \omega_o^2 x + a O(t)
\; .
\end{equation}

 In the presence of the coupling, if $a x(t)$ is sufficiently
small (an assumption we have to verify {\em a posteriori\/}),
 we can use linear response theory to calculate the action on $O$ of the 
oscillator:
\begin{equation}
O(t)=O_{\rm b}(t) + a \int_0^t dt'\,  R_O(t,t') x(t')
\; ,
\end{equation}
where $O_{\rm b}(t)$ is the fluctuating term
and where the response function $R_O$ is defined by
\begin{equation}
R_O(t,t')= \frac{\delta \langle  O \rangle (t)}{\delta a x(t')}
\; .
\end{equation}
We assume moreover that the average $\langle  O_{\rm b}(t) \rangle $
exists (we set it to zero by a suitable shift of $x$) and 
that the fluctuations of $O_{\rm b}$ (in the absence of coupling) 
are correlated as
\begin{equation}
\langle  O_{\rm b}(t)O_{\rm b}(t_{\rm w}) \rangle = C_O(t,t_{\rm w})\; ,
\label{bare}
\end{equation}
where $C_O(t,t_{\rm w})$ is a quantity of $O(N)$.

The equation for $x$ then reads
\begin{equation}
{\ddot x}= - \omega_o^2 x + a O_{\rm b}(t) + a^2  \int_0^t dt'\; R_O(t,t') x(t')
\; .
\label{osc}
\end{equation}
Thus the oscillator takes up energy from
the fluctuations of $O$, and dissipates it through
the response of the system.
Equation (\ref{osc}) is linear and easy to solve in the limit of small $a^2$ 
by Fourier-Laplace transform. One thus obtains the following results,
whose proof is sketched in Appendix \ref{APPthermo1}.

Consider first the case of stirred systems at stationarity, in
which both the correlation and the response are TTI.
The average potential energy of the oscillator reaches the limit
\begin{equation}
\frac{1}{2} \omega_o^2 \;  \langle x^2 \rangle  = 
 \frac{1}{2} \langle E_{\rm osc}\rangle =  
\frac{\omega_o   \,  \tilde C_O(\omega_o)}{ 
 2\chi_O''(\omega_o)}\;,
\end{equation}
after a time $\sim t_{\rm c}$ given by
\begin{equation}
t_{\rm c} = \frac{2 \omega_o}{a^2  \chi''(\omega_o) }.
\label{tc}
\end{equation}
We have defined
\begin{eqnarray}
\chi''(\omega) \equiv {\mbox{Im}} \int_0^\infty dt\, R(t)\, e^{i \omega t} 
\; ,
\nonumber\\
\tilde C(\omega) \equiv {\mbox{Re}} \int_{0}^{\infty} dt \, C(t)\, e^{i \omega t}
\; .
\end{eqnarray}

 If we now define the temperature $T_O(\omega)$ as follows:
\begin{equation}
  T_O(\omega) \equiv \langle E_{\rm osc} \rangle,
\end{equation}
we obtain
\begin{equation}
T_O(\omega) =   \frac{ \omega_o    \tilde C_O(\omega_o)  } 
{    \chi_O''(\omega_o)           }
\; .
\label{hohenberg}
\end{equation}
 This is precisely the temperature defined by Hohenberg
and Shraiman \cite{Hosh} for the case of weak turbulence (the spatial
dependence is encoded in $O(s)$). 

 Let us now turn to the case of relaxational dynamics where TTI is violated. Here we have
to take into account the time $t_{\rm w}$ at which the measurement is performed,
and consider $\omega_o^{-1} \ll t_{\rm c} \ll t_{\rm w}$. Then, a similar calculation
yields for the energy of the oscillator
\begin{equation}
  \frac{1}{2}  
 \omega_o^2 \;  \langle x^2\rangle_{t_{\rm w}} = 
\frac{1}{2} \langle E_{\rm osc}\rangle_{t_{\rm w}}=  
 \frac{\omega_o  \tilde C_O(\omega_o,t_{\rm w})}{ 
 2 \chi_O''(\omega_o,t_{\rm w})}
\label{result1}
\; ,
\end{equation}
where the average $\langle x^2\rangle_{t_{\rm w}}$ is taken on a comparatively
short time stretch after $t_{\rm w}$.
This definition of the frequency- and waiting-time-dependent
correlation $ C_O(\omega_o,t_{\rm w})$ and out-of-phase susceptibility
$\chi''_O(\omega_o,t_{\rm w})$ closely follows the actual experimental
procedure for their measure:
one considers a time window around $t_{\rm w}$ consisting of a few cycles
(so that phase and amplitude can be defined) and small enough respect
to $t_{\rm w}$ so that the measure is ``as local as possible in time''.
In fact, these two quantities are standard in the experimental
investigation of aging phenomena in spin-glasses \cite{agingSG,Bouchiat,Saclay}.

The natural definition of the frequency- and time-dependent temperature 
of $O$ is then
\begin{equation}
T_O(\omega_o,t_{\rm w}) \equiv 
 \frac{\omega_o  \,  \tilde C_O(\omega_o,t_{\rm w})}{ 
  \chi_O''(\omega_o,t_{\rm w})}
\; .
\label{effectT}
\end{equation}
If equilibrium is achieved, then 
the temperature is independent of $t_{\rm w}$, of the frequency $\omega_o$ and
of the observable $O$, and coincides with that of the heat bath. 
The index $O$ recalls that the effective temperature may depend 
on the observable.
The frequency-dependent temperature defined by either (\ref{hohenberg}) or
(\ref{effectT}) is compatible with the Fourier transformed expression
of the FDT violation factor (\ref{QFDT}) provided that it does not
vary too fast with $\omega_o$. We show later that this
is indeed the case for systems with slow dynamics.

For this definition we have chosen somehow arbitrarily an oscillator as
our thermometer. However, we  show in Appendix \ref{APPthermo2}
that the same role can be played by any small but macroscopic
thermometer, weakly coupled to the system. The role
of the characteristic
frequency $\omega_o$ is then played by the inverse of the typical response
time of the thermometer.

Now $ T_O(\omega,t_{\rm w})$  only deserves to be called a ``temperature''
if it controls the direction of heat flow.  
A first way to check whether this is the case is to consider an experiment
in which we connect the oscillator to an observable $O_1$ and let it
equilibrate at the temperature $T_{O_1}(\omega,t_{\rm w})$.
We then disconnect it and connect it to another observable $O_2$
and let it equilibrate at the temperature $T_{O_2}(\omega,t_{\rm w})$. 
This is not like a Maxwell demon, since the times of connection and
disconnection are unrelated to the microscopic behavior of the system.
The net result is that an amount of energy $T_{O_1}(\omega,t_{\rm w})-
T_{O_2}(\omega,t_{\rm w})$ has been transferred from the degrees
of freedom associated with $O_1$ to those associated with $O_2$:
therefore the flow goes from high to low temperatures.
This is an actual realization of the idea of
``touching two points of  the glass with a copper wire''
described in the Introduction.

This observation  also suggests a possible explanation of the fact that all
FDT violation factors which we know of are smaller than one:
if $T_O(\omega,t_{\rm w})$ were smaller than the temperature of the heat
bath, it could be possible, in principle, to extract energy from the bath
by connecting between it and our system a small Carnot engine.
The argument can be made sharper by considering a stationary
situation in a weakly stirred system: but to argue that this
situation would lead to a violation of the Second Principle one needs
to prove that the equilibration times of the Carnot engine are short
enough to make the power it produces larger than the dissipated one.

\section{Equilibration within a time-scale}
Before discussing the effective temperature as an equilibration 
factor we need to introduce some general features of the time-evolution
of systems with slow dynamics. We first define the time-correlation
scales and we then argue that, if the FDT violation factor
is well defined within a time scale, a single degree of
freedom thermalizes within that time scale at the corresponding
effective temperature. 

\subsection{Correlation and response scales}
 Systems having a long-time out-of-equilibrium dynamics tend to have
 different behaviours in different time-scales.
 Let us start by describing them for long-time dynamics in the  purely
 relaxational case. In Appendix \ref{APPscales} we give a formal definition 
of correlation scale, following Ref.~\cite{Cuku2}. 
 
 Consider first, as an example, the dynamics of  domain growth \cite{Bray}
 for the Ising model at low but non-zero temperatures.
 The autocorrelation function $C(t,t_{\rm w})=\frac{1}{N} \sum_i s_i(t)s_i(t_{\rm w})$
 exhibits two regimes:
 \begin{itemize}
\item
At long times $t,t_{\rm w}$, such that $t-t_{\rm w}\ll t_{\rm w}$, the correlation function 
 shows a fast decay from $1$ (at equal times) to $m^2$, where $m$
is the magnetization.
 This regime describes the fast relaxation of the spins within the bulk of
 each domain.
\item At long and well separated times ($t-t_{\rm w}\sim t_{\rm w}$),
 the correlation behaves as a function of $L(t_{\rm w})/L(t)$,
where $L(t)$ is some measure of the typical domain size at time $t$.
\end{itemize} 
We refer to these two scales as ``quasiequilibrium'' and ``coarsening'' 
(or ``aging''), respectively.  
 After a given time, the correlation rapidly decays to a plateau value $m^2$, and
the speed with which it falls below that value  becomes smaller and smaller
as $t_{\rm w}$ grows.

  This example helps to stress the fact that 
  different scales are defined as a function of 
both times (in this case $t-t_{\rm w}$ finite, 
  $L(t_{\rm w})/L(t)$ finite) and are well separated
 only in the limit where both times are large.

  In Fourier space, this separation of scales is achieved by
considering several frequencies $\omega$  and
 increasing times $t_{\rm w}$. Then, if we consider $\omega\sim {\rm const.}$ for increasing $t_{\rm w}$, we probe the quasiequilibrium scale, 
while if we want to probe the aging or coarsening
scale we have to 
consider smaller and smaller $\omega$, keeping $\omega L(t_{\rm w})/L'(t_{\rm w})\sim 
{\mbox{constant}}$.

If we now consider two frequencies $\omega_1$ and $\omega_2$,
and keep 
\begin{equation}
\frac{\omega_1}{\omega_2}\sim{\rm const.}\;,\label{scales}
\end{equation}
we shall probe, as $t_{\rm w}\to\infty$, the {\em same\/} scale:
it will be the quasiequilibrium scale if they both remain finite,
or the coarsening (aging) one if we keep 
$\omega_{1,2}L(t_{\rm w})/L'(t_{\rm w})\sim {\rm const.}$

These considerations can be generalized to other systems with
slow dynamics. In general there may be more than two relevant
scales \cite{So,Cuku1,Cuku2,Frme}, for example
 $\omega = {\rm const.}$,  $\omega t_{\rm w}^{1/2}={\rm const.}$,
  $\omega t_{\rm w}= {\rm const.}$, etc.
 In any case, the condition  for looking into the same scale
via two successions of frequencies remains Eq.~(\ref{scales}).

In Fig.~\ref{FIGp3ctwa1} we show the numerical solution of our test model,
defined in Appendix \ref{APPmodel}. The auto-correlation function  $C(t,t_{\rm w})$ 
is plotted vs.\ the time difference $t-t_{\rm w}$ for several 
waiting times in log-log scale. These plots (which are standard  in the 
 MonteCarlo simulations of  spin-glasses \cite{Ri}), show (i) 
that the system is out of equilibrium,
since we have an explicit dependence
on $t_{\rm w}$, and (ii) that there are at least two time-correlation scales. 
For short time differences $t-t_{\rm w}$ the decay is fast, the 
auto-correlation function is TTI  and it falls from 1 to $C^{EA}$. 
It then decays further from $C^{EA}$ to zero, 
more and more slowly as the waiting time
increases.
The quantity $C^{EA}$ is known in the 
language of spin-glasses as the Edwards-Anderson order
parameter and in Mode Coupling Theory (MCT) as
the non-ergodicity parameter: it measures the strength of the fast 
correlations ($C^{EA} = m^2$ in domain growth).
Quasiequilibrium and aging regimes
are experimentally observed in
real spin-glasses and polymer glasses\cite{struik,Saclay}.

 Let us now turn to a similar analysis for driven systems in the limit 
of weak driving energy $D \rightarrow 0$. In that case,
even if the system reaches a TTI regime provided we
wait long enough (the amount of waiting
increases when the stirring rate $D$ decreases),
it sometimes happens that some correlations and responses 
acquire a non-trivial low frequency behaviour in the limit $D \rightarrow 0$
\cite{Crsom}.
For example, in Fig.~\ref{FIGp3ctwalpha} we show the same plot as in Fig.~\ref{FIGp3ctwa1},  i.e.,
 $C(t-t_{\rm w})$  vs.\ $t-t_{\rm w}$,
for different, but small, stirring rates. We see
that there are at least two time scales: one, for short time
differences, where the 
correlation decays rapidly from 1 to $C^{EA}$ and one, for long time 
differences, in which it slowly decays from $C^{EA}$ to zero. 
The smaller the stirring, the slower the decay of $C$ from $C^{EA}$ to zero. 

It is useful to consider, in these cases, frequencies that go to zero as
some function of $D$.
The condition that two sequences of
frequencies $\omega_1,\omega_2$ correspond to the same scale now reads
\begin{equation}
\lim_{D \rightarrow 0} \frac{\omega_1}{\omega_2} =  {\mbox{const.}}
\; .
\label{scales1}
\end{equation}

\begin{figure}
\centerline{\epsfxsize=10cm
\epsffile{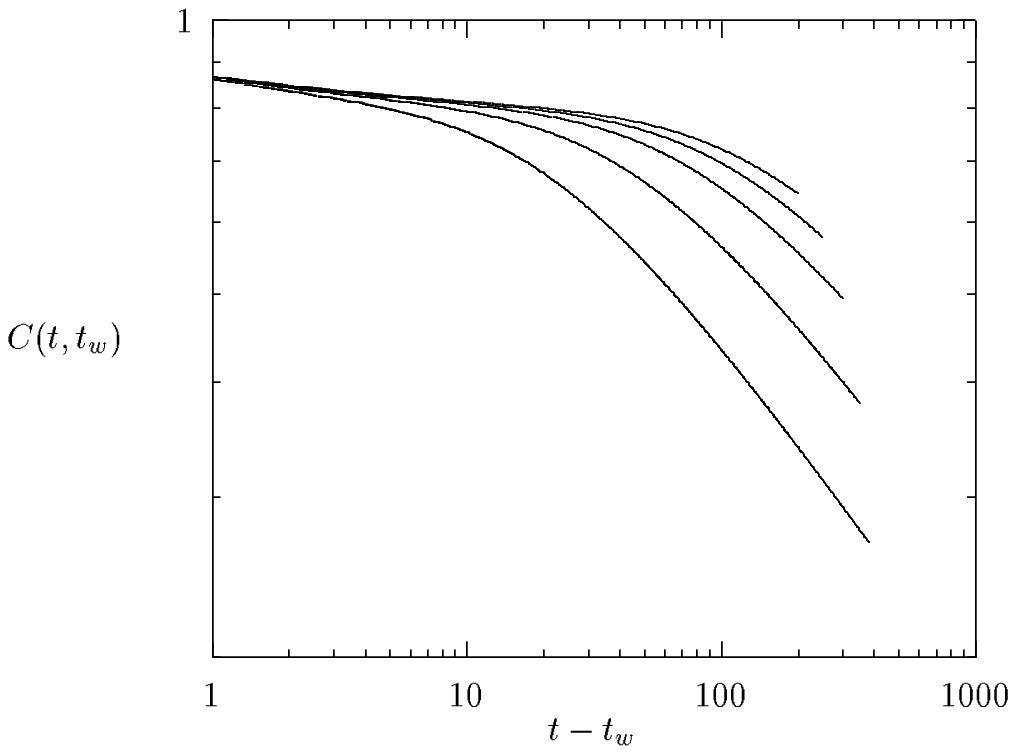}
}
\caption{The correlation function $C(t,t_{\rm w})$  vs.\ $t-t_{\rm w}$.
for the same model as in Fig.~\ref{FIGp3chitfinite}. 
From bottom to  top $t_{\rm w}=20,\;50,\;100,\;150,\;200$.
The correlation decays rapidly from 1 to $C^{EA} \sim 0.73$ and 
then more slowly from $C^{EA}$
to $0$. This second decay becomes slower and slower as $t_{\rm w}$ increases.
}
\label{FIGp3ctwa1}
\end{figure}
\begin{figure}
\centerline{\epsfxsize=10cm
\epsffile{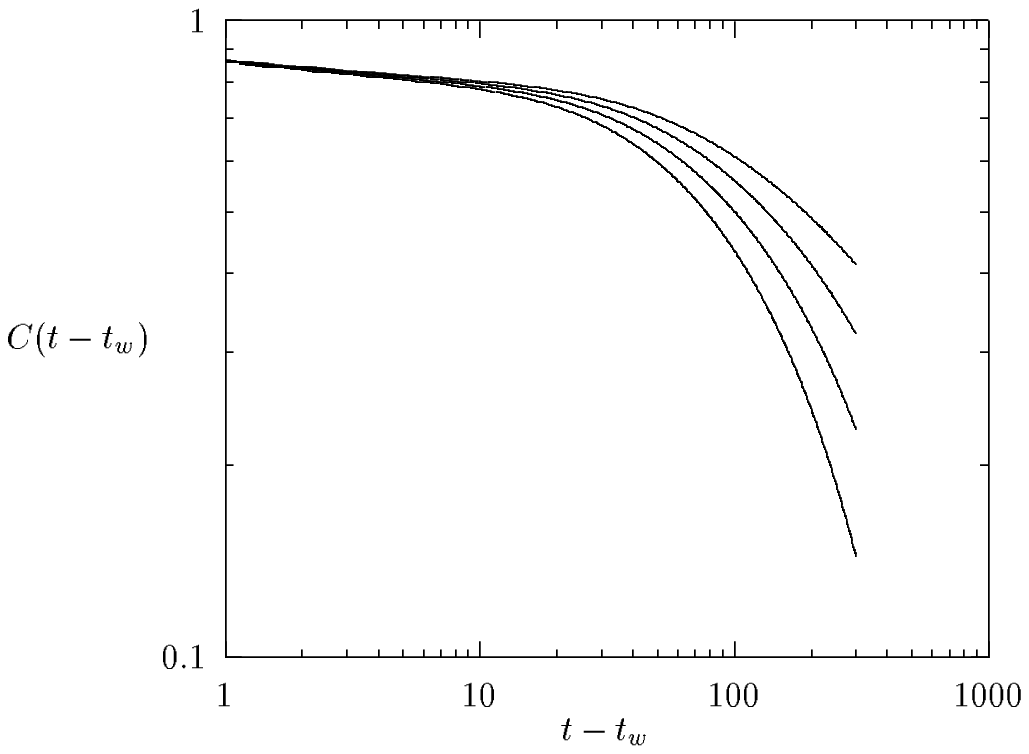}
}
\caption{The correlation function $C(t-t_{\rm w})$  
vs.\ $t-t_{\rm w}$ for the same model as in Fig.~\ref{FIGp3chialpha}.  
From  bottom to top, 
$D=0.1,\;0.075,\;0.05,\;0.025$. The correlation decays rapidly from
1 to $C^{EA} \sim 0.73$ and then more slowly from $C^{EA}$
to $0$. This second decay becomes slower and slower as $D$ decreases.
\label{FIGp3ctwalpha}
}
\end{figure}
\subsection{Thermalization criterion for a single degree of freedom
within a scale}
Let us now consider a system at a given waiting time $t_{\rm w}$ or
stirring rate $D$. Heat flows tend to zero as $t_{\rm w}\to\infty$
or $D\to 0$ respectively. It is the reasonable
to assume that, in these limits, the {\em effective temperatures\/}
associated with any given observable $O$ on a given time scale
tend to equalize, provided that their limit is finite.
We thus have, for example in the case of relaxational
dynamics, for $\omega_1/\omega_2={\rm const.}$,
\begin{equation}
\lim_{ {t_{\rm w} \rightarrow \infty}\atop{
 \; \omega_1\rightarrow 0}} 
T_O(\omega_1,t_{\rm w})=\lim_{{t_{\rm w}\to\infty}\atop{\omega_2\to 0}}
T_O(\omega_2,t_{\rm w})\;.
\label{univ}
\end{equation}
Similarly, we expect that each fixed frequency sooner or later thermalizes with the bath:
\begin{equation}
\lim_{ \stackrel{t_{\rm w} \rightarrow \infty } {\omega={\mbox{ const.}}}} 
T_O(\omega,t_{\rm w})=T\;.
\label{univ1}
\end{equation}
In some important cases the effective temperatures defined by
eqs.~(\ref{hohenberg}),(\ref{effectT}) diverge in the limit $t_{\rm w}\to \infty$
($D\to 0$). In these cases the effective temperature $T_O(\omega,t_{\rm w})$
should diverge for the whole time scale. We shall dwell on this
problem in Sec.~C.

 Equations (\ref{univ}) and (\ref{univ1}) are trivially true in a system that reaches thermal equilibrium, where all effective temperatures
eventually reach the temperature of the reservoir.
 However, they also describe the situation in which smaller 
frequencies take longer to reach the temperature of the heat bath,
in such a way that at
each given (long) waiting  time there are low enough frequencies that have
a temperature substantially different (in all cases we know, higher) from that of the bath.
In particular, Eq.~(\ref{univ1}) allows us to answer
a question we asked at the beginning: 
if a piece of glass has been kept at
room temperature for several months, a thermometer
whose response time is of order of a few seconds would measure
the room temperature, but it would read a higher temperature
if its response time is of the order of weeks.

In fact, with the appropriate handling of time-scales \cite{Cuku2}
(see Appendix \ref{APPscales}), Eqs.\ (\ref{univ}) and (\ref{univ1})
make it possible to calculate the out-of-equilibrium relaxation
of mean-field spin-glasses, and also the low temperature generalization of
the  mode-coupling equations for one single mode.
We thus obtain a solvable
example where Eqs.~(\ref{univ}) and (\ref{univ1}) hold.

For the case of stirred systems, the $D\to 0$ limit plays the same role
as the $t_{\rm w}\to\infty$ limit in relaxational systems, provided
that the time scales are suitably redefined as in Eq.~(\ref{scales1}).
In particular, at each fixed value of $\omega$ one has 
$\lim_{D\to 0}T_O(\omega, D)=T$, and, for $\omega_1/\omega_2={\rm const.}$
one has
\begin{equation}
\lim_{{D\to 0}\atop{\omega_1\to 0}} T_O(\omega_1,D)=
\lim_{{D\to 0}\atop{\omega_2\to 0}} T_O(\omega_2,D)\;,
\end{equation}
provided that the limits are finite.

If one looks into a time scale within which the temperature is almost constant,
 one  can indifferently use a temperature
defined in terms of a frequency or a time. In the relaxational case one has
\begin{equation}
T_{O}(\omega, t_{\rm w}) = T_{O}(t, t_{\rm w})\;,
\end{equation}
provided that $t-t_{\rm w}\sim \omega^{-1}$. In the same situation one has
\begin{equation}
 R(t,t_{\rm w})= \frac{1}{ T_O (t, t_{\rm w}) }  \frac{\partial C(t,t_{\rm w})}{\partial t_{\rm w}}\;.
\label{TTT}
\end{equation}
Similar relations hold for the driven case.
If, on the contrary, one considers values of frequency or time for which
the temperature defined by Eq.~(\ref{TTT}) is not constant,
one cannot directly relate it with the reading of a thermometer coupled to
the system.

\subsection{ Coarsening and the  case of infinite effective temperature}
\begin{figure}
\centerline{\epsfxsize=10cm
\epsffile{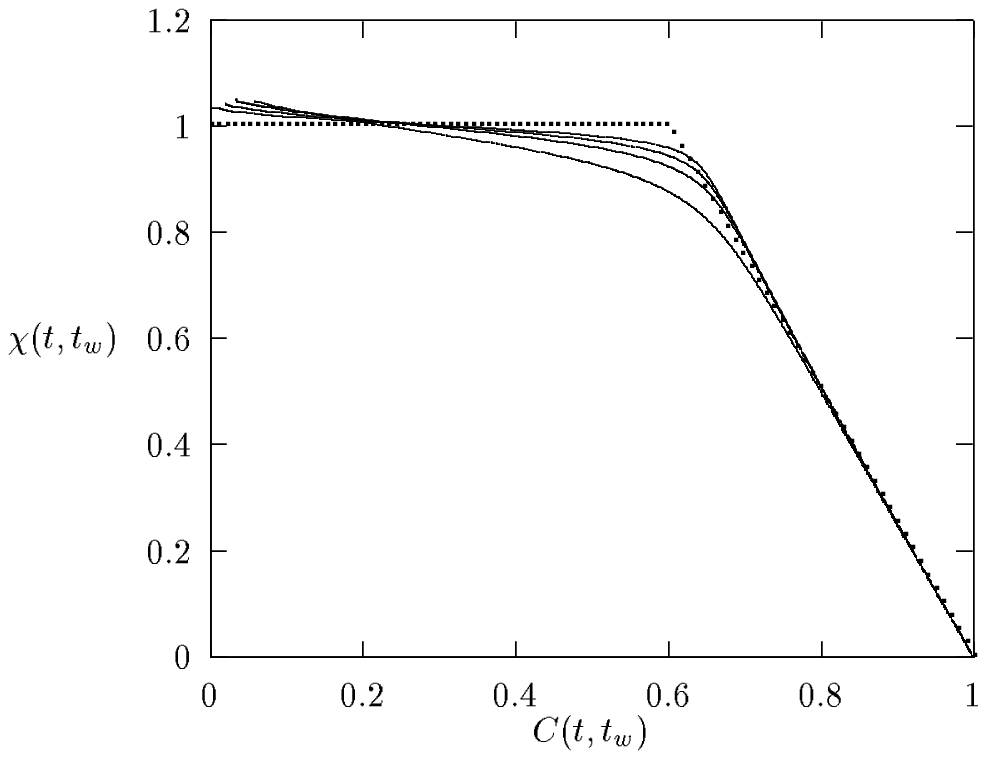}
}
\caption{The susceptibility $\chi(t,t_{\rm w})$ vs.\ the auto-correlation function $C(t,t_{\rm w})$
for the purely relaxational  $p=2$ spherical model 
(equivalent to the $O(N)$, $N \rightarrow \infty$ ferromagnetic
coarsening in $d=3$). The dots represent the analytical 
solution  when $t_{\rm w} \to \infty$. The total time
$t$ is equal to 20, 50, 100, 200 ($t_{\rm w}>t/4$ throughout).}
\label{FIGp2chitfinite}
\end{figure}

An important physical situation which deserves a
special discussion is that of domain growth.
In this case the effective temperature (associated, say, with the total magnetization)
in the ``coarsening scale'', $\omega L(t_{\rm w})/L'(t_{\rm w})={\mbox{const.}}$,
tends to infinity as $t_{\rm w}\to\infty$. 
The curve $\chi$ (magnetic susceptibility) vs.\ $C$ 
(magnetization correlation) looks like in Fig.~\ref{FIGp2chitfinite}.
This figure shows the results for
a model that is equivalent to the $O(N)$ ferromagnetic coarsening in three dimensions.
The integrated response  $\chi$ becomes flat,
as $t_{\rm w} \rightarrow \infty$,
 in the aging regime.
 In other words, the long-term
memory {\em tends to disappear}. 

Because experimental measures of aging are 
in general related to the response, this 
kind of system is sometimes referred to as exhibiting 
aging ``in the correlations'' (correlations are not TTI)
but not in the response.
One should stress, however, that the divergence of the effective temperature
in the aging time scale
can be extremely slow: for example, 
in the Fisher-Huse model for spin-glasses \cite{Fihu} 
$T_O(\omega,t_{\rm w})$ grows like a power of $\log t_{\rm w}$.

\begin{figure}
\centerline{\epsfxsize=10cm
\epsffile{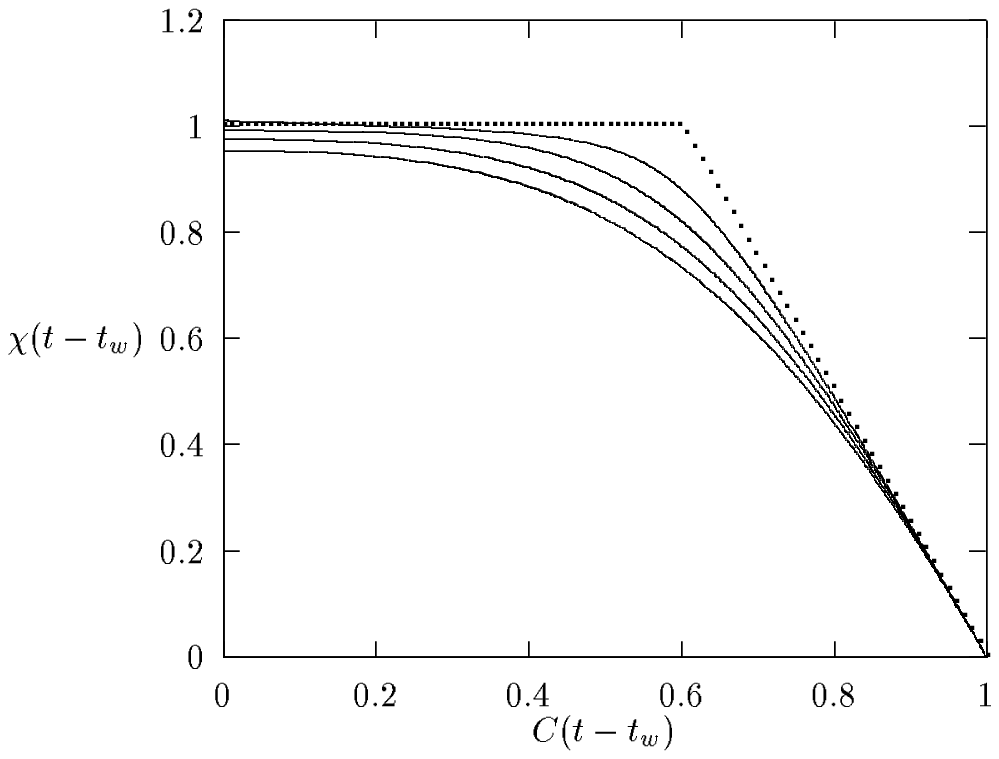}
}
\caption{The susceptibility $\chi(t-t_{\rm w})$ vs.\ the auto-correlation function $C(t-t_{\rm w})$
for the $p=2$ spherical asymmetric model (``stirred''  $O(N)$, $N \rightarrow \infty$ ferromagnetic
coarsening in $d=3$)
 for different levels of asymmetry (stirring):
from bottom to top,  $D=0.8,\; 0.6,\;0.4,\;0.2$. The dots represent the analytical 
solution in the limit of zero asymmetry. 
}
\label{FIGp2chialpha}
\end{figure}

 The fact that the effective temperature in the aging regime tends to infinity means that
 if we measure it at  $\omega$ small and fixed,
 the temperature $T(\omega,t_{\rm w})$ grows 
with $t_{\rm w}$ (while we are still probing the aging regime) and then starts falling, finally reaching
the temperature of the bath. 
If we repeat the experiment with a smaller $\omega$,
the overall behaviour is the same, but the highest temperature reached  will be higher; and so on 
without an upper limit.
The situation is similar, but experimentally more subtle,
for ``stirred'' systems for which the effective
temperatures diverge as $D\to 0$.
In such cases we measure a system that has a 
given (small) driving $D$.
If the system is such that for $D \rightarrow 0$ 
the effective temperature of all but the fastest scale
tend to infinity, we may never realize that this 
is the case in an experimental situation,
as long as we are not able to repeat
 the experiment with smaller and smaller stirring.

We show one such case in Fig.~\ref{FIGp2chialpha}, which represents
the same coarsening problem as before,
but made stationary by a ``stirring'' term in the equation of motion proportional to $D$.
If one had performed an experiment on such a system for a fixed $D$, 
one would have observed
high effective temperatures for separated times, and no evidence of thermalization.
Only by letting $D$ take smaller and smaller values one can
notice that in fact 
these temperatures tend to infinity.

\subsection{Systems with finite effective temperatures}
In mean-field models
one can close the Schwinger-Dyson equations into a 
set of dynamical equations involving only 
the correlation and response functions. 
This allows us to solve them analytically for large times, 
and also to obtain their full solution, 
via a numerical integration, starting from a pair of initial 
times $t_{\rm w}=t=0$. We thus obtain the curve $\chi(C)$  
by integrating the response function and plotting it vs.\
$C$. Two cases in which there are only two time-correlation scales 
are  the test model \cite{Cuku1} considered here (cf. Appendix \ref{APPmodel})
and the case
of a particle moving in an infinite-dimensional
random medium with short-range correlations \cite{Cule}. Both the analytical 
and the numerical results exhibit thermalization  within 
the aging regime.

In the Sherrington-Kirkpatrick spin-glass model there is a 
full hierarchy of time scales and effective temperatures,
 a fact also confirmed 
numerically \cite{Fe} and by Monte-Carlo simulations \cite{Cuku2}. 
Another model with infinite many scales and with a full  hierarchy of effective temperatures 
is that of  a particle moving in an infinite-dimensional
random medium with long-range correlations \cite{Frme,Cule}.

In Ref.~\cite{Frri}, the Monte-Carlo simulation of 
the ``realistic'' 3D Edwards-Anderson model for a spin-glasses 
was used to obtain the
 $\chi(C)$ curves, which seem to  approach a 
non-trivial curve for increasing $t_{\rm w}$. This suggests
the existence of a hierarchy of time-correlation scales. 

The ``trap model'' for spin-glass dynamics \cite{Bode}  violates Eq.(\ref{univ}) 
when an unusual choice of a parameter ($\alpha$) is made. However, it is 
difficult to interpret the model, with this choice of parameter, 
as a phenomenological model stemming from a reasonable microscopic 
dynamics.

\section{Thermalization of different degrees of freedom within a time scale}
 As we remarked in Section III, a {\em bona fide\/}
temperature should control heat flow and thermalization. 
It is the aim 
of this Section to show that this is 
indeed the case for the effective temperature
we have defined.  
If this were not the case, it would be possible to use our
small oscillator to transfer heat from some degrees of freedom
to others: in other words, by decorating our copper wire with
a suitable frequency filter, we would observe heat flowing through
it when it touches the two end of the glass.

We shall argue that if different degrees
of freedom {\em effectively interact\/} on a given time
scale, then they thermalize on that scale. A useful---though
not universal---criterion for ``effective interaction''  
is  that their mutual response
 (the response of one of the degrees of freedom to an
oscillating field of the given frequency  
conjugate to the other degree of freedom) is of the same 
order as their self-response on that time scale. 

The argument is essentially the same  both 
for  relaxational systems that have evolved for a long time, 
or in stationary driven systems
in the limit of small driving energy.
We shall thus focus on the relaxational case only.

We emphasize again that the thermalization
of different degrees of freedom is well defined only
in the limit of vanishing heat flow,
i.e., long waiting times or vanishing stirring
rates respectively.
This is witnessed by the appearance of one (or more) well
separated plateaus in the decay of the two-time correlation functions.

Let us consider $n$ modes (labelled by $a,b=1,\ldots,n$).
One can write, in general, some Schwinger-Dyson equations
for their correlations ${\sf C}=(C_{ab})$ and
responses ${\sf R}=(R_{ab})$:
\begin{mathletters}
\label{eq21}
\begin{eqnarray}
{\partial C_{ab}(t,t_{\rm w}) \over \partial t}
&=&
 - \sum_{c} \mu_{ac}(t) \, C_{cb}(t,t_{\rm w}) + 2 T \, R_{ab}(t_{\rm w},t) +
\nonumber\\
& &
+\sum_{c} \int_0^{t_{\rm w}} dt'' \; D_{ac}(t,t'') \; R_{cb}(t_{\rm w},t'')
+\sum_{c} \int_0^t dt'' \; \Sigma_{ac}(t,t'') \; C_{cb}(t'',t_{\rm w})\;,
\label{eq21a}
\\
{\partial R_{ab}(t,t_{\rm w}) \over \partial t}
&=&
 -\sum_c\mu_{ac}(t) \, R_{cb}(t,t_{\rm w}) +  \delta(t-t_{\rm w})\;\delta_{ab} +
 \sum_{c}  \int_{t_{\rm w}}^t dt'' \;
 \Sigma_{ac}(t,t'') \; R_{cb}(t'',t_{\rm w})\;.
\label{dyson}
\end{eqnarray}
\end{mathletters}
As they  stand, Eqs.~(\ref{eq21}) are just a way 
to hide our difficulties under the $\Sigma$, $D$ carpet.
Several approximations 
extensively used in the literature amount to various approximations 
of the kernels $\Sigma$ and $D$.

In equilibrium, the symmetries \cite{Bocukume} 
of the original problem allow us to write:
\begin{mathletters}
\label{FDTT5}
\begin{eqnarray}
\Sigma_{ab}(t-t_{\rm w})&= &\frac{1}{T} 
\frac{ \partial D_{ab}}{\partial t_{\rm w}} (t-t_{\rm w})\;,\label{FDT4}\\
R_{ab}(t-t_{\rm w})&= &\frac{1}{T} 
\frac{ \partial C_{ab}}{\partial t_{\rm w}} (t-t_{\rm w})
\; .
\label{FDT5}
\end{eqnarray}
\end{mathletters}
If we now make for $\Sigma$ and $D$ 
the approximation that are ordinary fuctions
(instead of general functionals) of the correlations and the responses, we obtain the
 Mode-Coupling Approximation (MCA)
\begin{equation}
D_{ab}({\sf C})=F_{ab}({\sf C}) 
\; ,
\qquad
\Sigma_{ab}({\sf C})=\sum_{c,d} F_{ab,cd}({\sf C}) R_{cd}
\; ,
\label{mctoff1}
\end{equation}
where there is a model-dependent function $F(\sf q)$ such that 
\begin{equation}
 F_{ab}({\sf q})=\frac{\partial F}{\partial q_{ab}};\qquad
 F_{ab,cd}({\sf q})=\frac{\partial^2 F}{\partial q_{ab}\partial q_{cd}}
\; .
\label{mctoff}
\end{equation}
%where $F$ is a given function determined by the model.

If the system equilibrates, one recovers the usual form of the Mode
Coupling Theory (MCT) that
is applied for supercooled liquids  \cite{Go}. 
It will be instructive to recall how equilibrium is reached
within the framework of eq.~(\ref{mctoff}).
One recalls
\begin{itemize}
\item
The FDT conditions (\ref{FDTT5});
\item
Time translational invariance: functions of one time are just 
constant and functions of
two times depend upon time differences only.
\item
Reciprocity: $C_{ab}(t-t_{\rm w})=C_{ba}(t-t_{\rm w})$.
\end{itemize}
Putting in this information in  Eqs.~(\ref{eq21})
and (\ref{mctoff}) one obtains
\begin{eqnarray}
{\partial C_{ab}(t-t_{\rm w}) \over \partial t}
&=&
 - \sum_{c} \mu_{ac} \, C_{cb}(t-t_{\rm w}) + 
  \frac{1}{T} \sum_{c} \left[
D_{ac}(0)
 \; C_{cb}(t_{\rm w}-t_{\rm w}) - D_{ac}^{\infty} C_{cb}^{\infty} 
\right]\nonumber \\
&&+ \frac{1}{T} \sum_{c} \int_{t_{\rm w}}^t dt''
 \;  D_{ac} (t-t'') \; \frac{ \partial C_{cb}(t''-t_{\rm w})} {\partial t''}
\; ,
\label{mcton}
\end{eqnarray}
where $D_{ac}^{\infty} , C_{cb}^{\infty}$ 
stand for limits of widely separated times,  i.e.,
\begin{equation}
C_{ac}^{\infty} \equiv \lim_{t \rightarrow \infty } C_{ac}(t)
\; ,
\qquad
D_{ac}^{\infty} \equiv \lim_{t \rightarrow \infty } D_{ac}(t)
\; .
\end{equation}
This is the usual equilibrium MCT equation \cite{Go}. 
The equation for the response (\ref{dyson}) 
with the same equilibrium assumptions becomes 
the time-derivative (divided by $T$) of Eq.(\ref{eq21a}), as it should. 

In a case in which the system is unable to reach
thermal equilibrium, like the low temperature phase of spin glasses,
the solution of Eqs.~(\ref{eq21})
will exhibit several time scales.
The question as to how many scales one has to consider in order to close
the dynamical equations can be answered for each model 
unambiguously by the construction in
Ref.~\cite{Cuku2}, suitably generalized to several modes.

Here,  for simplicity, we assume that there are only two relevant time
scales (as in the coarsening example of
the last section). We discuss later a model system
that acts as an explicit example of this situation. 
 We propose an ansatz for the long-time asymptotics, and then
verify that it closes the equations \cite{Cuku1}.
We assume (and later verify) that all two-time 
functions can be separated in two regimes:
\begin{itemize}
\item
Finite time differences with respect to the 
long waiting time, i.e.,
 $t-t_{\rm w} $ finite and positive, and  $t_{\rm w} \to \infty$.
In this regime TTI and FDT hold.
\item
 Aging regime, corresponding to long and  widely separated times,
i.e., $t\sim t_{\rm w}\to\infty$.
In this regime neither TTI nor FDT hold.
\end{itemize}

For finite time differences and a  long waiting time $t_{\rm w}$ 
we have:
\begin{mathletters}
\label{sepa}
\begin{eqnarray}
C_{ab}^{FDT} (t-t_{\rm w}) &\equiv& \lim_{t_{\rm w}\to\infty} C_{ab}(t,t_{\rm w})
\; , \\ 
 R_{ab}^{FDT} (t-t_{\rm w})  &\equiv& \lim_{t_{\rm w}\to\infty} R_{ab}(t,t_{\rm w})
\; ,
\end{eqnarray}
\end{mathletters}
with
\begin{eqnarray}
 R_{ab}^{FDT} (t-t_{\rm w})& = & \frac{1}{T} \frac{\partial 
C_{ab}^{FDT}}{\partial t_{\rm w}} (t-t_{\rm w})\;,\\
C^{EA}_{ab} &\equiv &\lim_{t-t_{\rm w} \rightarrow \infty} \lim_{t_{\rm w} \rightarrow \infty}  C_{ab}^{FDT}(t-t_{\rm w})\; .
\end{eqnarray}

The aging regime is defined as the time-domain 
in which the correlations fall (more and more slowly) below
$C^{EA}_{ab}$. For these times we denote
\begin{equation}
C_{ab} (t,t_{\rm w}) = \tilde{C}_{ab} (t,t_{\rm w}) \; , \qquad 
R_{ab} (t,t_{\rm w}) = \tilde{R}_{ab} (t,t_{\rm w}) 
\label{sepaaging}
\; .
\end{equation}

The separation  (\ref{sepa}) induces within the MCA a similar separation for $\Sigma$ and $D$ (cfr.\
Eqs.~(\ref{mctoff1}) and  (\ref{mctoff})), namely, for close times,
\begin{eqnarray}
D_{ab}^{FDT} (t-t_{\rm w})  &=& \lim_{t_{\rm w}\to\infty} D_{ab}(t,t_{\rm w})
\; , \nonumber \\ 
\Sigma_{ab}^{FDT} (t-t_{\rm w})  &=& \lim_{t_{\rm w}\to\infty} \Sigma_{ab}(t,t_{\rm w})\; ,
\label{sepa1}
\end{eqnarray}
where FDT holds:
\begin{equation}
 \Sigma_{ab}^{FDT} (t-t_{\rm w}) = \frac{1}{T} \frac{\partial D_{ab}^{FDT}}{\partial t_{\rm w}} (t-t_{\rm w})
\; .
\end{equation}
Again, we can define
\begin{equation}
\lim_{t-t_{\rm w} \rightarrow \infty} 
\lim_{t_{\rm w} \rightarrow \infty}  D_{ab}^{FDT}(t-t_{\rm w}) 
\equiv D^{EA}_{ab}\; ,
\end{equation}
and mark with the tilde the aging part of the kernels:
\begin{equation}
D_{ab} (t,t_{\rm w}) =\tilde{D}_{ab} (t,t_{\rm w}) \; ,    \qquad    
\Sigma_{ab} (t,t_{\rm w}) =\tilde{\Sigma}_{ab} (t,t_{\rm w})
\; .
\label{sepa1aging}
\end{equation}

In order to close the dynamical equations, we make an ansatz 
for the aging parts 
$\tilde{C}_{ab} (t,t_{\rm w})$ and $ \tilde{R}_{ab} (t,t_{\rm w})$. 
For a problem in which the correlations  vary only within  two time scales,
 the natural generalization of the solution in Ref.~\cite{Cuku1}
is
\begin{mathletters}
\label{ansatz}
\begin{eqnarray}
\tilde{C}_{ab}(t,t_{\rm w})&=& \tilde{C}_{ab}(h_{ab}(t_{\rm w})/h_{ab}(t)) 
\; ,
\\
\tilde{R}_{ab}(t,t_{\rm w})&=& \frac{X_{ab}}{T} \frac{\partial  \tilde{C}_{ab}}{\partial t_{\rm w}}(t,t_{\rm w})
\; ,
\end{eqnarray}
\end{mathletters}
where the $X_{ab}$ are constants and the $h_{ab}(t)$  are functions
 to be determined from
the dynamical equations. Note that the derivative in (\ref{ansatz}) is taken
with respect to the earliest time.

Remarkably, it turns out that one can close the equations 
with two different types of ansatz for the long-time
aging behaviour. In Appendix \ref{APPtwosyst} we show how this is done. 
In terms of the effective temperatures 
 the meaning of these two possibilities is:
\begin{enumerate}
\item {\it Thermalized aging  regime.}
 The effective temperatures associated with the observables $O_1$, $O_2$
 are  equal to each other for frequencies and waiting times
 in the aging regime:
they are not necessarily equal to the temperature of the bath.
At higher frequencies, they both coincide with the one of the bath.
\begin{equation}
\begin{array}{llll}
T_1(\omega,t_{\rm w})=T_2(\omega,t_{\rm w})=T\;,\quad&
t_{\rm w} \rightarrow \infty\;,\quad&C_{ab}>C_{ab}^{EA}\;,\quad&
 {\mbox{quasieq.}}\\
T_1(\omega,t_{\rm w})=T_2(\omega,t_{\rm w}) \neq T\;,\quad
&\omega  \rightarrow 0\;, \;\;t_{\rm w} \to \infty \;,\quad &
C_{ab}<C_{ab}^{EA}\;,\quad&
\mbox{aging.} 
\end{array}
\end{equation}

Not surprisingly, in this  case we find that $O_1$ and $O_2$ are
 strongly coupled ({\em also in the aging regime\/}) in the sense that
the mutual responses 
\begin{equation}
\tilde{R}_{12}(t,t_{\rm w})= 
\frac{X}{T} \frac{\partial  \tilde{C}_{12}}{\partial t_{\rm w}} \; ,\qquad\tilde{R}_{21}(t,t_{\rm w})= \frac{X}{T} \frac{\partial  \tilde{C}_{21}}{\partial t_{\rm w}}\;,
\end{equation}
where $X > 0$, are of the the same order of the self response functions.

\item {\it Unthermalized aging  regime.}
 The effective temperatures associated with the observables $O_1$, $O_2$ 
 for combinations of 
frequencies and waiting times corresponding to the aging regime are
neither equal to each other nor to that of the bath, while for higher frequencies
they both coincide with the one of the bath.

\begin{equation}
\begin{array}{llll}
T_1(\omega,t_{\rm w})=T_2(\omega,t_{\rm w})=T\;,\quad&
t_{\rm w} \rightarrow \infty\;,\quad &
 C_{ab}>C_{ab}^{EA}\;,\quad & {\mbox{quasieq.}}\\ 
T_1(\omega,t_{\rm w})\neq T_2(\omega,t_{\rm w}) \neq T\;,\quad & 
\omega  \rightarrow 0\;, \;\; t_{\rm w} \to \infty \;,
\quad &C_{ab}<C_{ab}^{EA}\;,&
\mbox{aging. } 
\end{array}
\end{equation}
In this case, 
$O_1$ and $O_2$ are effectively uncoupled ({\em in the aging regime}),
 in the sense that:
\begin{equation}
\tilde{R}_{12}(t,t_{\rm w}) = 
\frac{X_{12}}{T} \frac{\partial  \tilde{C}_{12}}{\partial t_{\rm w}}\;,
\qquad\tilde{R}_{21}(t,t_{\rm w}) = \frac{X_{21}}{T} \frac{\partial  
\tilde{C}_{21}}{\partial t_{\rm w}}\;,
\end{equation}
where
\begin{equation}
 X_{12},X_{21} \; \rightarrow 0
\;.
\end{equation}
\end{enumerate}

We have not found any other way of closing the equations \cite{fank}. 
We shall show below, 
in a particular case which
can be numerically solved, that indeed
either case (1) or case (2) take place.
Let us remark that a solution with 
$X_{12}=X_{21}\neq 0$, $X_{11}=X_{22}=X\neq 0$
and $X\neq X_{12}$ is not compatble with the interpretation of $X$
as an inverse temperature. We do not find such a solution in our
test models and believe that it is not
realizable in general.

The considerations of this section  can also 
be made in the limit of small stirring.

\begin{figure}
\centerline{\epsfxsize=10cm
\epsffile{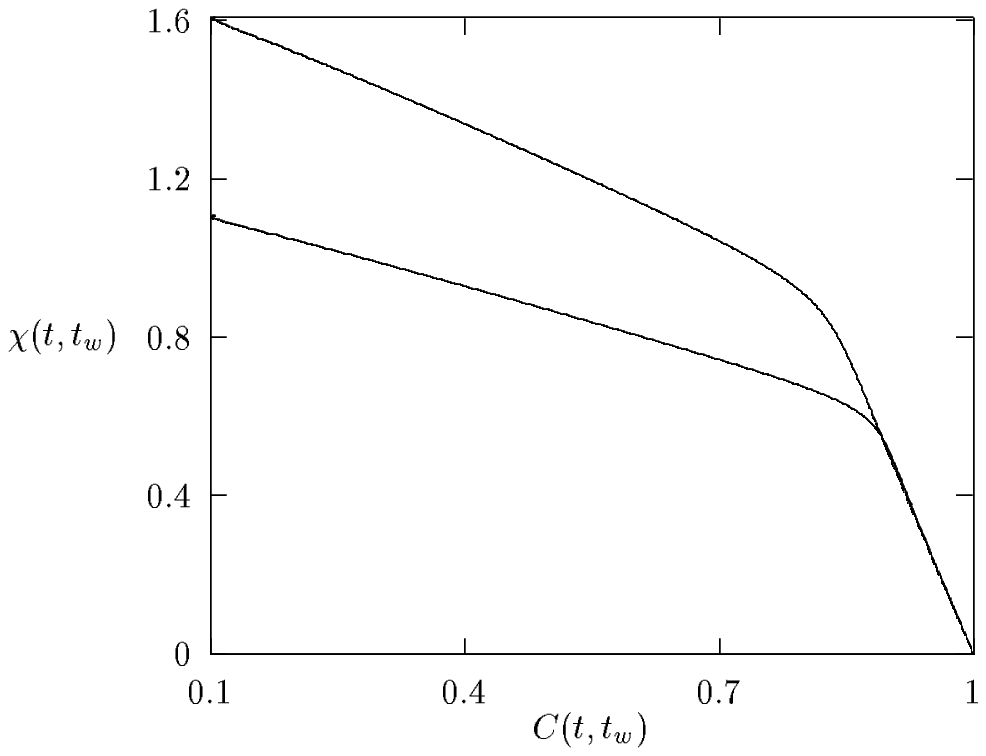}
}
\caption{The susceptibility $\chi_{11}(t,t_{\rm w})$ and $\chi_{22}(t,t_{\rm w})$ vs.\ the corresponding 
auto-correlation functions $C_{11}(t,t_{\rm w})$ and $C_{22}(t,t_{\rm w})$
for the   {\it uncoupled}, aging systems. The slopes of the curves, i.e.,
the FDT violation factors and hence the effective temperatures
are different. This corresponds to the {\it unthermalized\/} case.
}
\label{FIG2syste0}
\end{figure}

\begin{figure}
\centerline{\epsfxsize=10cm
\epsffile{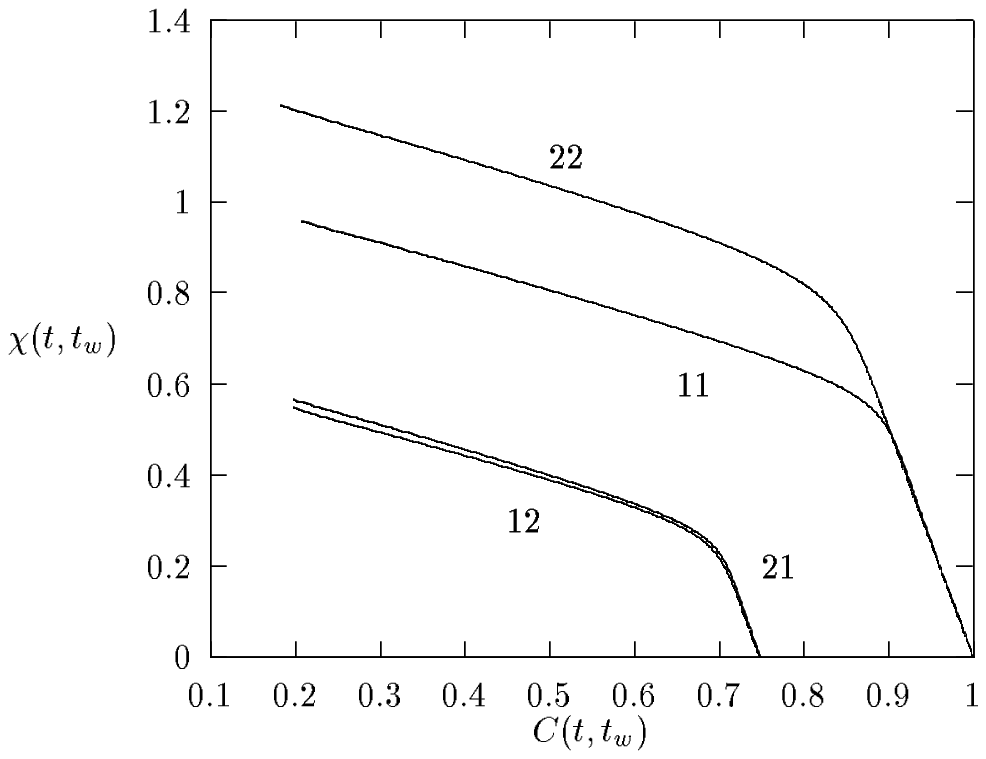}
}
\caption{The susceptibilities $\chi_{ij}(t,t_{\rm w})$ vs.\ the correlation functions $C_{ij}(t,t_{\rm w})$
for the two aging systems of Fig.~\ref{FIGp3chitfinite}, this time  {\it weakly\/} coupled.
 The FDT violation factors $X_{ij}(C_{ij})$ are almost parallel: thermalization is almost complete.
}
\label{FIGtwosysth05}
\end{figure}
\begin{figure}
\centerline{\epsfxsize=10cm
\epsffile{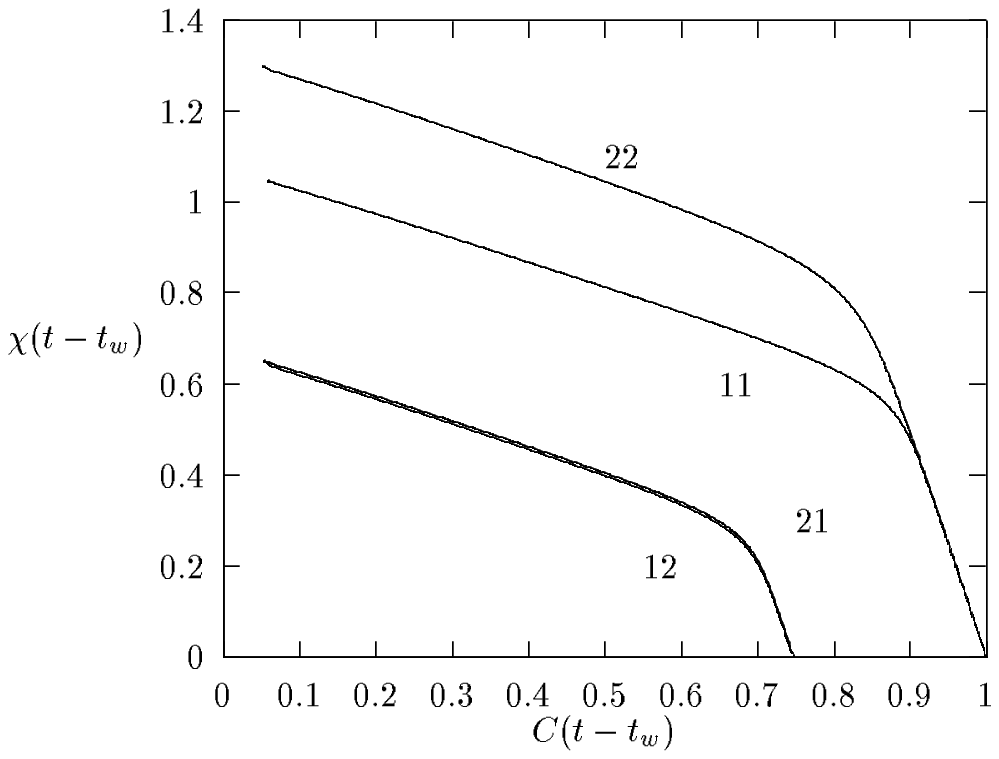}
}
\caption{The susceptibilities $\chi_{ij}(t-t_{\rm w})$ vs.\ the correlation functions $C_{ij}(t-t_{\rm w})$
for two slightly driven system  ($D=0.1$), {\it weakly\/} coupled.
 The FDT violation factors $X_{ij}(C_{ij})$ 
given by the slopes of these curves are now the same: 
after a short transient corresponding to finite
times all the curves become parallel.  The aging regimes have {\it thermalized\/}.
}
\label{FIGtwosystasym}
\end{figure}

\subsection{An explicit  model with thermalization}

 In order to test in a particular example
 that these  asymptotic solutions are not
only consistent, but are in fact reached, we solve numerically
the mode coupling equations with two coupled modes, with $F$ given by:
\begin{equation}
F({\bf q})= q_{11}^p + K^2 q_{22}^p
\; .
\label{pspin}
\end{equation}
We impose normalization at equal times of the autocorrelation of both modes:
\begin{equation}
C_{11}(t,t)= C_{22}(t,t)=1
\; .
\end{equation}

As usual with the mode coupling equations \cite{Kr,Amro,Fiso,Frhe,Bocukume}, one can find a disordered mean field model
whose dynamics is exactly given by these equations. This is the  model  defined  in Appendix \ref{APPmodel}.

Figures \ref{FIG2syste0} and \ref{FIGtwosysth05} show the numerical solution of the exact system of 
coupled  integro-differential  equations whose  large-time asymptotic  
can be analytically obtained as  in Appendix \ref{APPtwosyst}. In Fig.~\ref{FIG2syste0}  and 
\ref{FIGtwosysth05} we plot $\chi(C)$ for two cases. In Fig.~\ref{FIG2syste0}
we consider two  uncoupled systems that evolve from the initial 
condition $C_{11}(0,0)=C_{22}(0,0)=1$ and $C_{12}(0,0)=C_{21}(0,0)=0$.
We see that the systems evolve independently and $X_{11} \neq X_{22}$ 
while $X_{12}= X_{21}=0$ 
(unthermalized case). In 
Fig.~\ref{FIGtwosysth05} we consider the same global system but 
now including a  weak coupling between the two individual systems. 
Clearly, after a short transient associated to short times, all curves
$\chi_{ij}(C_{ij})$ get parallel. 
The systems aging-regime temperatures for the two subsystems 
have  become equal (thermalized case). 

It is interesting to note that, in this example,
Onsager's reciprocity relations hold:
$R_{12}=R_{21}$, $C_{12}=C_{21}$. However, one can imagine
situations in which they do not hold separately,
but where however $X_{12}=X_{21}$.

For comparison, we show in 
Fig.~\ref{FIGtwosystasym} the corresponding plot for two weakly coupled systems 
in the limit of small stirring.

\subsection{Many scales}

 Let us briefly discuss what would happen in the presence of many
scales, each one with its own temperature. 

Consider, for instance, two observables $O_1$ and $O_2$, their associated
 autocorrelation functions
$C_{11}(t,t_{\rm w})$, $C_{22}(t,t_{\rm w})$,  their integrated
 self-responses $\chi_{11}(t,t_{\rm w})$, $\chi_{22}(t,t_{\rm w})$
and the effective temperatures $T_1(C_{11})$, $T_2(C_{22})$.
We can plot  $T_1(C_{11})$ vs. $C_{11}$
 and  $T_2(C_{22})$ vs. $C_{22}$.
These two plots need not be the same,
 even if all scales are thermalized.
Consider now a parametric plot of $C_{11}(t,t_{\rm w})$  vs.
 $C_{22}(t,t_{\rm w})$, in the limit of very
large $t_{\rm w}$: it defines a function $C_{11}= {\cal H}(C_{22})$ that
allows one to calculate $C_{11}$  for large $t,t_{\rm w}$, given 
 $C_{22}(t,t_{\rm w})$.

The condition for thermalization in every time-scale is then
 that the curve $T_2(C_{22})$
coincides with the curve $T_1({\cal H}(C_{22}))$, both considered
 as functions of $C_{22}$:
\begin{equation}
T_1({\cal H}(C_{22}))=T_2(C_{22}) \;.
\label{superthermo}
\end{equation}
The deviation of $T_1({\cal H}(C_{22}))$
 from $T_2(C_{22})$ is a measure of the degree in which the two
 observables are not thermalised.

Actually, (\ref{superthermo}) was obtained as an ansatz for the
 dynamics of a manifold in a random medium within the
Hartree approximation \cite{Cukule},
where the role of different observables is played by 
the displacements at different spatial
wavelengths:  
\begin{eqnarray}
C_k(t,t')={\cal H}_k(C_{k=0}) 
\; ,
\nonumber \\
T_{k}({\cal H}_k(C_{k=0}))=T_{k=0}(C_{k=0})
\; .
\label{superthermo1}
\end{eqnarray}
In this case, the modes at different $k$ thermalize at the same effective temperature,
although their mutual responses vanish, as implied by translational symmetry in mean.

\section{Comparison with other effective temperatures}

The  idea of ``fictive temperatures''~\cite{To} $T_{\rm f}$
in  glasses goes back to the 
'40s  and it has developed since
\cite{Nara,fictrev,agingglass}. 
Here we recall it briefly, for the sake of comparison
with the  effective temperature that we have discussed.

When cooling a liquid  the 
time needed to establish equilibrium grows and,
eventually, the structural change cannot keep
pace with the rate of cooling: the system falls out of equilibrium 
and enters the glass transition region. It is then said that 
``the structure is frozen'' at a temperature  characterized by a 
 fictive temperature $T_{\rm f}$. The fictive temperature
defined in terms of different quantities of interest,  e.g.,
the enthalpy, the thermal expansion coefficient, etc., do not 
necessarily coincide. Furthermore, 
it has been experimentally observed  that glasses with the same 
fictive temperatures arrived at through different preparation paths
may  have different molecular structures. The fictive temperature is
hence a phenomenological convenience and should not be 
associated with a definite molecular structure \cite{agingglass}.

The fictive temperature is a function 
of the temperature of the bath $T$.   
At high temperature,  when the sample
is in the liquid phase,  $T_{\rm f}=T$. When the liquid enters the transition
range $T_{\rm f}$ departs from $T$ and $T_{\rm f} >T$; and, finally, deep below the 
transition range, where the relaxation is fully stopped, $T_{\rm f} \to 
T_{\rm g}$.
The detailed bath-temperature dependence of the fictive temperature
in the region of interest is usually  
expressed by~\cite{fictrev,agingglass}
\begin{equation}
\dot T_{\rm f}  = -\frac{T_{\rm f}-T}{\tau(T_{\rm f}, T)} \; ,
\label{tfdyn}
\; ,
\end{equation}
where the characteristic time  $\tau(T_{\rm f}, T)$ depends both upon $T_{\rm f}(t)$ and upon the 
thermal history of the sample given by $T(t)$. At equilibrium $T_{\rm f}=T$ 
and $\dot T_{\rm f}=0$.
This nonlinear differential equation 
determines $T_{\rm f}(t)$ once one {\it chooses\/} $\tau(T_{\rm f}, T)$.
A commonly used expression is the Narayanaswamy-Moynihan
equation
\begin{equation}
\tau(T_{\rm f}, T) = \tau_o \; \exp\left[\frac{x A}{T} + 
\frac{(1-x) A}{T_{\rm f}} \right]
\; ,
\label{taudef}
\end{equation}
where $\tau_o$, $A$ and $x \in [0,1]$ are some constants. All the
information about the dynamics of the system enters into $T_{\rm f}$ through 
these constants. 

In order to obtain the time relaxation of the quantity of interest,
the picture is completed by proposing a given relaxation function,
like the stretched exponential, the Davidson-Cole function, etc., 
and by introducing the fictive temperature through 
the characteristic time  $\tau(T_{\rm f}, T)$. See Ref.~\cite{agingglass} for an 
extensive discussion about the 
applications of $T_{\rm f}$ to the description of experimental data.

One may wonder  whether  is a relation between this {\it fictive\/} temperature 
of glass phenomenology 
and the {\it effective\/} temperature we have been discussing in this paper.

First of all, one notes that the fictive temperatures are defined through 
the relaxation  of the observables, unlike the one we consider here 
which are defined 
rather in terms of fluctuations and responses.
Both temperatures may depend upon the observable and
 upon the thermal history of the sample.

The dependence of the effective 
temperature defined in Eq.~(\ref{FIGp3ctwa1}) upon $T$ depends on the model.
In all cases FDT holds in the high temperature 
phase and $T_O(\omega,t_{\rm w})=T$, the effective temperature 
is equal to the bath temperature.
When entering the low temperature phase, the temperature
dependence of the effective temperature observed at
fixed low frequency depends on the model. In certain simple mean-field
 (or low-temperature mode-coupling) models whose dynamics 
following a  quench into the
glass phase can be solved, one can compute $T_O(\omega,t_{\rm w})$ explicitly. 
Three different behaviours 
are found:
\begin{itemize}
\item In the 
simple model we have been using as a test example in the previous 
sections \cite{Cuku1}, $T_O(\omega,t_{\rm w})$ starts from $T_O(\omega,t_{\rm w})=T_{\rm g}$ at $T=T_{\rm g}$
and slightly {\em increases}  when the 
bath temperature decreases below $T_{\rm g}$.
One would  instead expect a fictive temperature to 
remain stuck to  $T_{\rm g}$ in a (mean-field) model in which the glass
 transition is sharp.
\item There are other mean-field models such as the 
Sherrington-Kirkpatrick spin-glass model
in which there are infinitely many different effective
temperatures. The lowest aging-regime temperature appears discontinuously ($T_O(\omega,t_{\rm w}) > T_{\rm g}$)
as one crosses $T_{\rm g}$ and can be shown \cite{Cuku2}
to {\em decrease\/} with decreasing temperature.
Another example of this kind 
is the model of a particle moving in a random potential 
with long-range correlations \cite{Frme,Cule}.
\item In all cases we know of domain growth models \cite{Bray}  one obtains
$X(C)  \to 0$ for $t_{\rm w} \to \infty$, 
in the whole low temperature phase. Hence $T_O(\omega,t_{\rm w}) \to \infty$ in the
aging (coarsening) regime for all heat bath temperatures below the ordering
transition. This behaviour also holds for
certain extremely simple disordered systems such as the 
spherical Sherrington-Kirkpatrick spin-glass \cite{p2}, the 
toy domain growth model we used in this paper.
\end{itemize}

It is important to remark that these simple examples
 have the (sometimes unrealistic) feature that nothing depends permanently 
upon the cooling procedure. One expects, however, that more refined models that go beyond
the mean-field approximation will capture a cooling rate dependence that will also become
manifest in the 
effective temperature.

Another attempt to identify and relate a  microscopic effective temperature
to the fictive temperature of  glasses was put forward
by Baschnagel, Binder and Wittmann \cite{Babiwi} in the context of a 
lattice model for
polymer melts. They have pointed out  that 
in this model the usual FDT relation between the specific heat with the energy fluctuations
is broken at low temperatures and have tried to identify the 
FDT violation factor with an expression they propose 
for an {\it internal temperature}. 
 The internal temperature defined in Ref.~\cite{Babiwi}
has, however,  the unpleasant property of 
not reducing to the bath temperature in the high temperature phase.

In Ref.~\cite{Frrit} Franz and Ritort have also discussed the possibility of
relating the FDT ratio with an effective  temperature, in  the particular 
case of the Backgammonn model. They have compared the 
the value of $X(t,t_{\rm w})$ for finite times $t,t_{\rm w}$ with the temperature arising
from an  adiabatic approximation they used to  solve the model. 
The result is negative, in the sense that $T/X(t,t_{\rm w})$ does not coincide 
with the ``adiabatic temperature''.

These examples suggest that the phenomenological fictive temperatures act essentially as parameters
for describing an out-of-equilibrium ``equation of state'' while the effective temperature we have discussed 
plays a role closer to the thermodynamical one.

\section{Discussion, experimental perspectives and conclusions}

Although our discussion has been biased by the models we can solve at present,
we feel that the concept of the effective temperature that we discussed should be relevant
for many systems {\em with small energy flows}.
Indeed,   the key observation we make is that this temperature actually controls thermalization
and heat flows within a time-scale.
Therefore, the effective temperature can be a starting point for the investigation of the
thermology and, hopefully, the thermodynamics of out-of-equilibrium systems with small energy
flows.

In order to use this idea as a guiding concept for the planning of
experiments, one has to take some care: obviously, waiting times can
be large but not infinite in experiments, and stirring rates can be small but not
infinitesimal. Moreover, the models that we have explicitly discussed do not
exhibit dependence on the cooling history of the sample. Now, one would expect
in general the effective temperature to depend on $t$, $t_{\rm w}$ and,
 e.g., on the
cooling rate. One can hope to catch this aspect in more refined models.

One should also pay attention to the equilibration times. 
The effective temperature measured by a small oscillator is related to the FDT violation
factor provided that the equilibration time $t_{\rm c}$ (Eq.(\ref{tc}))  remains much smaller than $t_{\rm w}$.
It sometimes happens that an observable $O$ which should thermalize at an effective 
temperature $T_O(\omega, t_{\rm w})$ exhibits so small fluctuations at this frequency that 
this time becomes unbearably long. This is the case, for instance, of high $k$ modes 
in the aging time scale for a manifold in a random potential. Therefore, although all $k$ modes 
nominally thermalize at the same effective temperature, only low enough $k$ modes can be effectively
used to measure it. How low $k$ must be will depend on the time scale one is looking at.

Several numerical and real experiments in structural glasses can be envisaged. For example, 
one can compare density-fluctuations and compressibility in different length scales in order
to check if they are equilibrated within a time-scale.  Since the low-temperature extension of 
the MCT makes definite predictions on the value of the first non-trivial effective 
temperature appearing as one crosses the transition (see Section VI), this provides a concrete 
ground for a experimental testing of MCT.

\begin{figure}
\label{coil}
\centerline{\epsfxsize=10cm
\epsffile{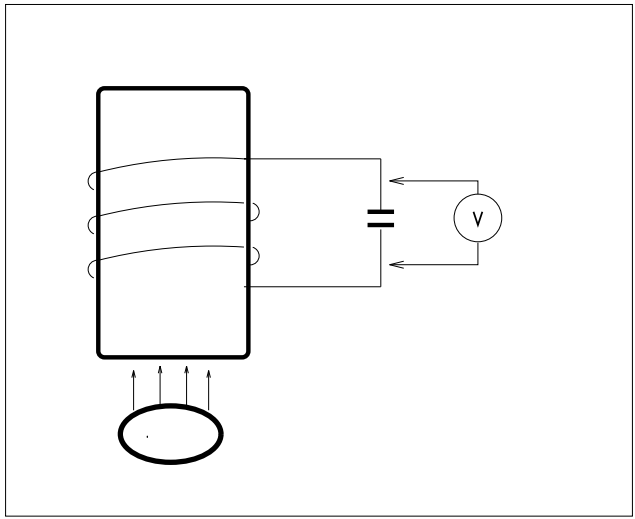}
}
\caption{An effective temperature measurement for a magnetic system. The coil
is wound around the sample, which is in contact with the bath. Coil and capacitor
have zero resistance
}
\label{FIGdibu}
\end{figure}

We close by 
considering the following (slightly {\em Gedanken}) 
spin-glass experiment of Fig. \ref{FIGdibu}:
  Currents are induced in the coil by the magnetization noise of the 
spin-glass, which is in contact with a heat bath. Apart from
the interaction with the sample, the L-C circuit of coil and 
capacitor is without losses.
 This is exactly a realization of the oscillator as a thermometer of 
Section II. 
 From what we know from the time-scales of real spin-glasses \cite{agingSG,Saclay}, 
if the time after
the quench is of the order of 10 minutes, and the period of the L-C circuit
 is of the order of the
second, we are probing (at least partially) the aging regime: the temperature 
 (defined as
the average energy of the capacitor) should be different from the bath 
temperature. We believe that it would be interesting to return to the magnetization 
noise experiments \cite{Bouchiat} with the purpose of measuring the effective
temperature: this would give us, for instance,  useful insights into the nature of the 
spin-glass transition.

\section*{acknowledgements}
We are specially indebted to  Andrea Baldassarri and Gilles Tarjus since this work initiated
from discussions with them. We also thank Gilles Tarjus for
introducing us to the old ideas of ``fictive temperatures'' used in glass phenomenology.
We acknowledge useful discussions with 
 S. Ciliberto, 
D. Dean, P. Le Doussal, M. M\'ezard, R. Monasson and P. Viot.
L. P. fondly remembers discussions with the late S.-K. Ma, who
nailed into his head the concept expressed by
the first sentence of this paper, and acknowledges the
support of a Chaire Joliot de l'ESPCI.
L.F.C. thanks for hospitality the Service de Physique Th\'eorique at Saclay, 
where this work was started, and the European Union
for financial support thorugh the contract ERBCHRXCT920069.

\appendix
\section{The test model}
\label{APPmodel}

Throughout this paper we have used as a test model a 
spherical disordered model with $p$-spin interactions.
The model has been introduced in its purely relaxational version 
by Crisanti and Sommers \cite{Crhoso} 
as a simple spin-glass model with several advantages, in particular, that exact dynamical equations can 
 be written for it in the thermodynamic limit. As shown by Franz and Hertz \cite{Frhe}, 
these dynamical equations are also those obtained from the MCA to the 
Amit-Roginsky model \cite{Amro}.
When considered in full generality, the two-time dynamical equations correspond to the low temperature 
extension \cite{Frhe,Bocukume} of the simplest mode coupling theory for the supercooled liquid phase
proposed by Leutheusser
\cite{Le} and Bengtzelius, G\"otze and Sj\"olander \cite{Begosj,Go}. 
It has been also 
recently shown by Chandra {\it et al.} that this model is related to a mean-field 
approach to Josephson junction arrays \cite{ioffe}.

We consider a system of $N$ variables ${\bf s}=(s_1,\ldots,s_N)$, subject
to forces $F^J_i$ given by
\begin{equation}
F_i^J({\bf s})= \sum_{\{j_1,\ldots,j_{p-1}\}}
 J_{i}^{j_1\ldots j_{p-1}} s_{j_1} \ldots s_{ j_{p-1}},
\end{equation}
where the couplings are random Gaussian variables. For different sets of indices
 $\{i ,j_1,\ldots,j_{p-1}\}$ the $J$'s are uncorrelated, 
while for permutations of the same set
of indices they are correlated so that
\begin{equation}
\overline{F^J_i({\bf s'})F^J_j ({\bf s})}= \delta_{ij} f_1(q)+s_i s'_j f_2(q)/N,
\end{equation}
where  $q=({\bf s}\cdot{\bf s}')/N$. 
In the purely relaxational case, one has
$f_2(q)=f'_1(q)$. We take here
$f_2(q)=(1-D) f_1'(q)$, where $f_1(q)=p  q^{p-1}/2$. The 
couplings $ J_{i}^{j_1\ldots j_{p-1}}$ are symmetric under
the permutation $i \leftrightarrow j_k$ in the purely relaxational case
($D=0$).
On the other hand, if 
$ J_{i}^{jj_2\ldots j_{p-1}}$
and $ J_{j}^{i j_2\ldots j_{p-1}}$ are uncorrelated, one has $D=1$.

The dynamics is of the  Langevin type:
\begin{equation}
\dot{s_i}= -F^J_i({\bf s}) - \mu(t) s_i - h_i(t)+\eta_i(t),
\end{equation}
where $\eta$ is a white noise of variance $2T$,
 $\mu(t)$ is a Lagrange multiplier 
enforcing the spherical constraint $\sum_{i=1}^N s_i^2 =1$, and $h_i(t)$ 
is an external field
(usually set to zero), needed to define the response functions.

In Figs.~\ref{FIGp3chialpha} and \ref{FIGp3chitfinite} we plot the $\chi(C)$ curves for the asymmterical and symmetrical 
$p=3$ model, respectively. In Figs.~\ref{FIGp3ctwa1} and \ref{FIGp3ctwalpha} we plot the auto-correlation decays 
for the symmetrical and asymmetrical versions respectively.  Figures \ref{FIGp2chitfinite} and \ref{FIGp2chialpha} 
show the $\chi(C)$ curves for the $p=2$ version that is analogous to the $O(N)$ model 
in $D=3$ when $N \to \infty$. 

 In order to check thermalization in this  particular example
we consider the evolution of two such systems with spins $s$ and $\sigma$, 
 and uncorrelated realizations of disorder $J$ and $J'$ and thermal noise $\eta$ and $\eta'$,
respectively.
They are coupled via the term proportional to $\mu_{12}$ and $\mu_{21}$ in  
the Langevin equations
\begin{mathletters}
\label{eq:p2}
\begin{eqnarray}
\frac{\partial s_i}{\partial t} &=& - F^J_i(s) - \mu_{11}(t) \, s_i - \mu_{12} \sigma_i + h_i
+ \eta_i\; , \\
\frac{\partial \sigma_i}{\partial t} &=& -K F^{J'}_i(\sigma)- \mu_{22}(t)  \, \sigma_i - \mu_{21} s_i +h'_i
+  \eta'_i
 \label{p2bis}
\; .
\end{eqnarray}
\end{mathletters}
We set $\mu_{12}=\mu_{21}$, so that the coupling does
not contribute to the stirring. The coefficients $\mu_{ii}$ are 
the Lagrange multipliers for each system.
The factor $K$ is introduced to break the symmetry between the subsystems.
The correlations 
\begin{equation}
\begin{array}{ll}
C_{11}(t,t_{\rm w}) = \frac{1}{N} \sum_{i=1}^N  
\langle s_i(t)s_i(t_{\rm w}) \rangle \; ,  
\qquad
&C_{22}(t,t_{\rm w}) =\frac{1}{N} \sum_{i=1}^N \langle 
\sigma_i(t)\sigma_i(t_{\rm w})\rangle 
\; ,\\
C_{12}(t,t_{\rm w}) = \frac{1}{N} \sum_{i=1}^N 
\langle s_i(t) \sigma_i(t_{\rm w}) \rangle \; ,  \qquad
&C_{21}(t,t_{\rm w}) =\frac{1}{N} \sum_{i=1}^N \langle 
s_i(t)\sigma_i(t_{\rm w}) \rangle
\; ,
\end{array}
\end{equation}
and responses
\begin{equation}
\begin{array}{ll}
R_{11}(t,t_{\rm w}) = \frac{1}{N} \sum_{i=1}^N 
\delta   \langle s_i(t) \rangle / \delta h_i^s (t_{\rm w}) 
\; ,  \qquad
&R_{22}(t,t_{\rm w}) = \frac{1}{N} \sum_{i=1}^N 
\delta  \langle \sigma_i(t) \rangle /\delta h_i^\sigma(t_{\rm w}) 
\; ,
 \\
R_{12}(t,t_{\rm w}) = \frac{1}{N} \sum_{i=1}^N 
\delta  \langle s_i(t) \rangle  /\delta  
h_i^\sigma(t_{\rm w})   \; ,  \qquad
&R_{21}(t,t_{\rm w}) = \frac{1}{N} \sum_{i=1}^N 
\delta \langle s_i(t) \rangle   / \delta h_i^s (t_{\rm w}) \;, 
\end{array}
\end{equation}
precisely satisfy Eqs.~(\ref{eq21a})-(\ref{mctoff}) with $F$ given by (\ref{pspin}).

The  results for the effective temperatures obtained from 
the numerical integration
of the exact evolution equations
of these systems are shown in Figs.~\ref{FIG2syste0} and
\ref{FIGtwosysth05}. 
See main text for the discussion.

\section{Energy of the oscillator}
\label{APPthermo1}

In this Appendix we solve Eq.~(\ref{osc}), 
we compute $\frac{1}{2} \omega_o^2 \, \langle x^2(t) \rangle$,
the average potential energy of the oscillator.
We thus prove Eq.~(\ref{result1}) and its form
 Eq.~(\ref{hohenberg})  valid for the stationary case.

Let us define
\begin{mathletters}
\label{ass0}
\begin{eqnarray}
\chi(\omega,t) \;\exp(i \omega t) &\equiv & \int_0^t dt' \; R(t,t') \exp(i \omega t') 
\; ,
\label{ass1}
\\
C(\omega,t) \; \exp(i \omega t)  &\equiv& \int_0^t dt' \; C(t,t') \exp(i \omega t')  
\label{ass2} 
\; .
\end{eqnarray}
\end{mathletters}
If $\omega^{-1} \ll t$, we can assume that $ \chi(\omega,t)$ and $C(\omega,t)$ are 
functions that
vary slowly with $t$, thus defining 
a Fourier component that is ``local'' in time $t$.

In general 
\begin{equation}
\langle x^2(t) \rangle 
=
a^2 \int_o^t  dt' \int_0^t dt'' \; G(t,t') \, G(t,t'') \, C(t',t'')\;,
\label{eqx2}
\end{equation}
where $G(t,t')$ is the Green function for the oscillator plus the  term 
representing the response of the system and $C(t,t')$ is the system's
auto-correlation function. Using the definition in Eq.(\ref{ass1}) one can show 
that the damped oscillator's Green function reads
\begin{eqnarray}
G(\omega,t) 
&=&
\frac{1}{-\omega^2+\omega_o^2 -a^2 \chi''(\omega,t)}
\; ,
\\
G(t,t') 
&=&  
\exp\left(-\frac{t-t'}{t_{\rm c}(t)}\right) \sin(\omega_o(t-t')) \; \theta(t-t')
\; .
\end{eqnarray}
We have here replaced $\chi(t,\omega)$ by $\chi''(t,\omega)$
using the fact that $a^2 N \ll 1$. The characteristic time $t_{\rm c}(t)$ 
of the damped oscillator is 
given by 
\begin{equation}
t_{\rm c}(t) = \frac{2 \omega_o}{a^2  \, \chi''(\omega_o,t)}
\; .
\end{equation}

We can now study  Eq.(\ref{eqx2}) by using the above  expressions for $G(t,t')$.
After a simple change of variables, using causality and the fact that 
$G(t,t')$ decays exponentially as a function of time-differences,
Eq.(\ref{eqx2}) can be rewritten as
\begin{equation}
\langle x^2(t) \rangle 
=
a^2 \int_{-\infty}^\infty  \frac{d\omega}{2\pi} 
\int_{-\infty}^\infty  d\tau \int_{-\infty}^\infty d\tau' \; 
G(t,t-\tau) \, G(t,t-\tau') \, C(\omega,t-\tau) \, \exp(i\omega(\tau-\tau'))
\; .
\label{eqx22}
\end{equation}
The fast exponential decay of the Green function 
allows us to replace $C(\omega,t-\tau)$ by $C(\omega,t)$.
Thus,
\begin{eqnarray}
\langle x^2(t) \rangle 
&=&
a^2 \int_{\infty}^\infty \frac{d\omega}{2\pi} \; G(\omega,t) \, G(-\omega,t) \, C(\omega,t)
= a^2 \int_{\infty}^\infty \frac{d\omega}{2\pi} \;
\frac{ C(\omega,t)}{\chi''(\omega,t)-\chi''(-\omega,t)} \;
\nonumber\\
&&\times\left[ \frac{1}{\omega_o^2- \omega^2 -a^2 N \chi''(-\omega,t)} - 
       \frac{1}{\omega_o^2- \omega^2 -a^2 N \chi''(\omega,t)} 
\right]
\; .
\end{eqnarray}
This integral can be calculated by the method of residues. We can close the 
circuit on the upper complex half plane. Since $\chi''(\omega,t)$ is analytic 
in the upper half plane, the only singularities are the zeroes of the denominators. 
Assuming that $a^2 \chi''$ is small, they lie in the vicinity of $\omega=\pm \omega_o$. 
In fact one can check that only two poles penetrate inside the circuit 
of integration. 
We obtain therefore
\begin{equation}
\langle x^2(t) \rangle = \frac{2 \tilde C(\omega_o,t) }{
\omega_o\,\chi''(\omega_o,t)} 
\label{result}
\; .
\end{equation}

This is the general result for the temperature.
The particular result (\ref{hohenberg}) that holds for the stationary case
is recovered from Eq. (\ref{result}) by letting $\tilde C(t,\omega)$
and $\chi''(t,\omega)$ be independent of $t$. Thus, (\ref{result})
reduces to (\ref{hohenberg}). 

\section{Small but macroscopic thermometers}
\label{APPthermo2}

In this Appendix we show that the thermometric considerations
made in Section III do not  crucially depend on the choice
of an oscillator as a thermometer. We use here 
a small but {\it macroscopic} thermometer defined by the 
variables $y_i$, $i=1,\ldots,n$, and we couple it only to the observable 
$O(s)$ of the system through a degree of freedom $x(y)$. 

Our measurement procedure
is as follows: we first thermalize the thermometer with an auxiliary bath
at temperature $T^*$. We then disconnect it from the bath and we connect 
it to the system through $O$. If there is no flow of energy between thermometer and 
system, then we conclude that the measured temperature is $T^*$.   

The energy of the thermometer plus its coupling with the system is
\begin{equation}
H =  E(y) - a O(s) x(y)
\; .
\end{equation} 
The net power gain of the thermometer is then $\dot Q(t)$, given by
\begin{equation}
\dot Q
= 
a \langle \dot O x  \rangle 
=
a \left. \partial_{t'}
\langle x(t) O(t') \rangle  \right|_{t'\to t^-}
\; .
\end{equation}
We look for the  condition that ensures stationarity for the thermometer. 
The thermometer is characterized by a temperature-depedent 
correlation $C_x(t,t')=\langle x(t) x(t') \rangle$ and its 
associated response $R_x(t,t')$.  
Using linear response one has
\begin{mathletters}
\begin{eqnarray}
O(t) &=& O_{\rm b}(t) + a \int_0^t dt' \, R_O(t,t') x(t')
\; ,
\\ 
x(t) &=& x_b(t) + a \int_0^t dt' \, R_x(t,t') O(t')
\; .
\end{eqnarray}
\end{mathletters}
To leading order in $a$, $\langle O x  \rangle $ is given by
\begin{equation}
\langle O x  \rangle 
=
a \int_0^t dt'' \; \left( R_x(t,t'') \, C_O(t'',t') + R_O(t',t'') \, C_x(t,t'') \right)
\; .
\end{equation}
Assuming now that $T^*$ is such that the thermometer can be considered to be almost in
equilibrium (note that the coupling $a$ is small and that we have chosen a thermometer
that is not itself a glass!), we obtain
\begin{equation}
\dot Q = a^2 \int_0^t dt' \; R_x(t-t') \; \left( \frac{\partial C_O(t,t')}{\partial t'} - T^*
R_O(t,t') \right)
\; .
\end{equation}

The condition for having no flow is then that 
the average of the parenthesis in the integral is zero. The weight 
function for this average is $R_x(t-t')$ which contains the characteristic 
time of the thermometer.

\section{Definition of time-correlation scales}
\label{APPscales}

In this appendix we review briefly the 
definition of  correlation ``time-scale'' introduced in Ref.~\cite{Cuku2}
for a correlation function that depends nontrivially upon two times.

Given a correlation function $C(t,t_{\rm w})$, 
which we assume normalizable in the large-times limit
 $C(t,t) \rightarrow C_\infty>0$, we consider
three  increasing times 
$ t_1<t_2<t_3$, and the limit in which they all go
to infinity, but in a way to keep $C(t_3,t_2)=a$ and $C(t_2,t_1)=b$
constant. We thus define the limit
\begin{equation}
\lim_{ {t_1,t_2,t_3\to\infty}\atop{C(t_3,t_2)=a \; , \; C(t_2,t_1)=b } } 
C(t_3,t_1) \equiv f(a,b)
\; .
\end{equation}

The mere existence of the limit ``triangle relation'' $f$ has extremely strong consequences:
considering four times one can easily show that $f$ is associative
\begin{equation}
f(a,f(b,c))=f(f(a,b),c)
\; .
\label{2}
\end{equation}
The form of an associative function on the reals is very restricted, and a classification of all
possible forms can be made \cite{Cuku2}.

It is sometimes convenient to work with the ``inverse''$\bar f$  of $f$ defined as
\begin{equation}
f(a,b)=c \qquad \Rightarrow  \qquad {\bar f}(a,c)=b
\; .
\end{equation}

We can now define a correlation scale in the following way: given two values of the correlation
at large times 
$C(t,t_{\rm w})={\cal C}^1$ and $C(t,t_{\rm w}')={\cal C}^2$, $t>t_{\rm w}'>t_{\rm w}$ and ${\cal C}^1 < {\cal C}^2$,
they are in a different correlation scale if
\begin{equation}
{\bar f}({\cal C}^2,{\cal C}^1)= {\cal C}^1
\; ,
\end{equation}
and are in the same scale otherwise.
In other words, the time it takes  the system to achieve ${\cal C}^2$ 
is negligible with respect to the time it takes to achieve ${\cal C}^1$.

\section{Solution of the two coupled-modes equation}
\label{APPtwosyst}

 Using the separation (\ref{sepa})-(\ref{sepa1aging}) and the mode-coupling approximation 
(\ref{mctoff1})-(\ref{mctoff})
we obtain a similar separation for $D_{ab}$ and $\Sigma_{ab}$:
\begin{mathletters}
\label{sepa1app}
\begin{eqnarray}
D_{ab}^{FDT} (t-t_{\rm w})&=&F_{ab}({\sf C}^{FDT} (t-t_{\rm w})) 
\; ,\\
\Sigma_{ab}^{FDT}(t-t_{\rm w})&=&\sum_{c,d} F_{ab,cd}({\sf C}^{FDT} (t-t_{\rm w})) R_{cd}^{FDT}(t-t_{\rm w}) 
\; ,
\end{eqnarray}
\end{mathletters}
where ${\sf C}^{FDT}$ stands for the set $C_{ab}^{FDT}(t-t_{\rm w})$.
One similarly obtains, in the aging regime,
\begin{mathletters}
\label{sepa2}
\begin{eqnarray}
{\tilde D}_{ab} (t,t_{\rm w})&=&F_{ab}({\sf {\tilde C}} (t,t_{\rm w})) 
\; ,\\
{\tilde \Sigma}_{ab}(t,t_{\rm w})&=&\sum_{c,d} F_{ab,cd}({\sf{\tilde C}} (t,t_{\rm w})) {\tilde R}_{cd}(t,t_{\rm w}) 
\; .
\end{eqnarray}
\end{mathletters}
 We can now write two coupled sets of  equations, valid in the quasiequilibrium regime and 
in the aging regime, respectively.
 For $t-t_{\rm w}$ finite, and  large times  $t>t_{\rm w}$, we have:
\begin{eqnarray}
{\partial C^{FDT}_{ab}(t-t_{\rm w}) \over \partial t}
&=&
 - \sum_c \left[ \mu_{ac}^\infty + \frac{D_{ac}^{FDT}(0)}{T}   \right]
\, C^{FDT}_{cb}(t-t_{\rm w}) 
-
\frac{1}{T} \, \sum_c D_{ac}^{EA} C_{cb}^{EA} 
+
M_{ab}^\infty\nonumber \\
 & &
+ \frac{1}{T} \sum_{c} \int_{t_{\rm w}}^t dt''
 \;  D^{FDT}_{ac} (t-t'') \; \frac{ \partial C^{FDT}_{cb}(t''-t_{\rm w})} {\partial t_{\rm w}}
\; ,
\label{mctonapp}
\end{eqnarray}
where $\mu_{ac}^\infty \equiv \lim_{t\to\infty} \mu_{ac}(t)$ and
\begin{equation}
M_{ab}^\infty 
\equiv 
\sum_c \lim_{t\to\infty} \int_0^t dt'' \, \left[ {\tilde D}_{ac}(t,t'') 
\tilde R_{cb} (t,t'') 
+ \tilde \Sigma_{ac}(t,t'') \tilde C_{cb}(t,t'') \right]
\; .
\label{Fdtreg}
\end{equation}
In the aging regime, for $t>t_{\rm w}$ we have
\begin{mathletters}
\label{eq21aapp}
\begin{eqnarray}
\frac{\partial {\tilde C}_{ab} (t,t_{\rm w})}{\partial t}
&=&
 - \sum_c \left[ \mu_{ac}(t) + \frac{D^{FDT}_{ac}(0)-D_{ac}^{EA} }{T} \right] \, {\tilde C}_{cb}(t,t_{\rm w}) +
 \sum_c 
{\tilde D}_{ac}(t,t_{\rm w})  \;  \frac{C^{FDT}_{cb}(0)-C_{cb}^{EA} }{T}  
\nonumber\\
& &
+ \sum_{c} \int_0^{t_{\rm w}} dt'' \; {\tilde D}_{ac}(t,t'') \; {\tilde R}_{cb}(t_{\rm w},t'')
+\sum_{c} \int_0^t dt'' \; {\tilde \Sigma}_{ac}(t,t'') \; {\tilde C}_{cb}(t'',t_{\rm w})
\; ,
\\
\frac{\partial {\tilde R}_{ab}(t,t_{\rm w})}{\partial t}
&=&
 - \sum_{c} \left[ \mu_{ac}(t) + \frac{D^{FDT}_{ac}(0)-D_{ac}^{EA} }{T} \right] \, {\tilde R}_{cb}(t,t_{\rm w}) + 
 \sum_{c}  \int_{t_{\rm w}}^t dt'' \;
 {\tilde \Sigma}_{ac}(t,t'') \; {\tilde R}_{cb}(t'',t_{\rm w}) 
\nonumber \\
 & & 
+  \sum_{c} {\tilde \Sigma}_{ac}(t,t_{\rm w}) \;\; \frac{C^{FDT}_{cb}(0)-C_{cb}^{EA} }{T}
\; . 
\end{eqnarray}
\end{mathletters}
Equation (\ref{Fdtreg}), for given $M^\infty_{ab}$ 
is very similar to the high-temperature 
Mode-Coupling equations \cite{Go}, and
can be solved in the same way.
An asymptotic solution for the aging regime 
can be obtained  by using the  generalization to more
than one mode
of the ansatz in Ref.~\cite{Cuku1} (\ref{ansatz}).
The derivative terms in Eqs.~(\ref{eq21aapp}) can be then dropped 
provided that $X_{11}\neq 0$ and $X_{22}\neq 0$,
a fact to be verified {\em a posteriori}. 
We shall find  in this way two different solutions for  
Eqs.~(\ref{eq21aapp}).

In the unthermalized case $X_{11}$ and  $X_{22}$
are different from zero, and possibly different
from each other, while $X_{12}=X_{21}=0$.
It is then easy  to see that Eqs.~(\ref{eq21aapp}) 
become effectively uncoupled {\em in this regime}, and
can be solved  \cite{Cuku1} as two separated one-mode equations, 
with the ansatz
\begin{mathletters}
\label{ansatzunth}
\begin{eqnarray}
\tilde{C}_{aa}(t,t_{\rm w})&=& \tilde{C}_{aa}(h_{aa}(t_{\rm w})/h_{aa}(t))\;, \\
\tilde{R}_{aa}(t,t_{\rm w})&=& \frac{X_{aa}}{T} \frac{\partial  
\tilde{C}_{aa}}{\partial t_{\rm w}}(t,t_{\rm w})\;, \\
\tilde{C}_{12}(t,t_{\rm w})&=&\tilde{C}_{21}(t,t_{\rm w})=0\; .
\end{eqnarray}
\end{mathletters}

In the thermalized case, one assumes 
\begin{mathletters}
\begin{eqnarray}
&&X_{11}=X_{22}=X_{12}=X_{21}=X \neq0 \;,\\
&&h_{11}=h_{22}=h_{12}=h_{21}=h\;.
\end{eqnarray}
\end{mathletters}
Making the change of variables 
\begin{equation}
\lambda \equiv \frac{h(t_{\rm w})}{h(t)} \;,\qquad
\lambda' \equiv \frac{h(t'')}{h(t)}\;,
\end{equation}
one finds that only the dependence on $\lambda$
survives in the equations, and that they reduce to a set
of four (instead of eight)  consistent 
equations for the correlations (see \cite{Cuku1,Cule} for
a systematic approach).

In this way, both for the thermalized and  the unthermalized case one can obtain $X_{ab}$, $M_{ab}$
and $C^{EA}_{ab}$. One then  has  to check that  $X_{11}\neq 0$ and $X_{22}\neq 0$. If this is not
the case the equations become identities,
and one cannot anymore neglect the derivative term.
One has therefore to
use a more refined long time limit.
These values have to be substituted
in  (\ref{Fdtreg}), in order to complete the solution in both
regimes.

Let us remark here that the problem of selecting
the functions $h_{ab}$ remains open.
This is an asymptotic matching problem in a non-local equation, 
and does not appear to be easily solvable.

We have thus found the long-time limit of the 
correlations and responses. If 
there is more than one asymptotic solution (even modulo $h$), 
we don't know for the time being
which asymptotic form is selected by the unique solution
of the evolution equations, without resorting to explicit
numerical integration.

In short we have:
\begin{enumerate}
\item 
{\it Unthermalized aging regime.}
\begin{equation}
X_{12} \rightarrow 0 \;,\qquad X_{21}  \rightarrow 0\;,\qquad
X_{11} \neq X_{22} \; .
\end{equation}
We have therefore
\begin{equation} 
\tilde{R}_{11}(t,t_{\rm w})= \frac{X_{11}}{T} \frac{\partial  
\tilde{C}_{11}}{\partial t_{\rm w}} 
\; ,
\qquad
\tilde{R}_{22}(t,t_{\rm w})= \frac{X_{22}}{T} 
\frac{\partial  \tilde{C}_{22}}{\partial t_{\rm w}} 
\; ,
\end{equation}
and
\begin{equation} 
\tilde{\Sigma}_{11}(t,t_{\rm w})= \frac{X_{11}}{T} 
\frac{\partial  \tilde{D}_{11}}{\partial t_{\rm w}} 
\; ,
\qquad
\tilde{\Sigma}_{22}(t,t_{\rm w})= \frac{X_{22}}{T} 
\frac{\partial  \tilde{D}_{22}}{\partial t_{\rm w}} 
\; .
\end{equation}

\item {\it Thermalized aging regime.}
\begin{eqnarray}
X_{11}=X_{12}= & X_{21} & =X_{22} \equiv X 
\; ,	
\\ 
\tilde{\Sigma}_{ab}(t,t_{\rm w})&=& \frac{X}{T} \frac{\partial  \tilde{D}_{ab}}{\partial t_{\rm w}} 
\; ,
\qquad\forall \;\; a,b
\; .
\end{eqnarray}
where the aging time-scales ``lock in'', i.e.,
there is the {\em same\/} function $h(t)$ for all $a$, $b$, such that
\begin{equation}
\tilde{C}_{ab}(t,t_{\rm w})= \tilde{C}_{ab} (h(t_{\rm w})/h(t))
\; .
\label{lock}
\end{equation}
This property can also be stated by saying that as 
 $t,t_{\rm w} \rightarrow \infty$ a plot of $\tilde{C}_{11}(t,t_{\rm w})$ 
vs.\
 $\tilde{C}_{22}(t,t_{\rm w})$ yields a single smooth curve: i.e.,
that there is a function ${\cal H}(C)$ such that
\begin{equation}
\tilde{C}_{11}(t,t_{\rm w})= {\cal H}(\tilde{C}_{22}) (t,t_{\rm w})\;.
\end{equation}
\end{enumerate}


\begin{references}
\bibitem{Foot0}
In the case of stirred systems, it sometimes happens that the stationary
regime sets in in a time that grows with decreasing $D$. In these cases,
we shall take the observation time large enough to guarantee stationarity.
\bibitem{FootSo} To the best of our knowledge, the first to propose that
violations of FDT where relevant for glassy dynamics was Sompolinsky \cite{So},
albeit within a framework which is now believed not to represent purely
relaxational systems. In fact,
Sompolinsky's ``barrier crossing'' dynamics is 
not an equilibrium dynamics (it yields a
wrong value for the equilibrium correlation functions), and it is not an 
out-of-equilibrium 
dynamics either (it assumes TTI, which is only valid in equilibrium).
The dynamics
for times that are not very separated \cite{Sozi}, however,
can be well justified, and FDT holds in that regime of times.
See Refs.~\cite{Hojayo,Biyo,Mepavi} for a discussion.
\bibitem{FootHo} A closely related situation is when the parameters
describing the Hamiltonian are made to be varied slowly. This approach 
has been advocated by Horner \cite{Ho} as a tool to ``regularize''
(i.e., render stationary) the dynamics of spin-glasses, thus avoiding
having to deal with waiting-time dependences.
\bibitem{fank}An algebraically similar phenomenon occurs in the static
treatment of two coupled systems with replicas, when an ansatz is made
as in Ref.~\cite{kky}. In that context, however, the meaning of the ansatz
is still uncertain. We still lack a probabilistic interpretation in
terms of pure states, as in the ordinary Parisi ansatz.
\end{references}
\end{document}